\documentclass[12pt]{article}
\usepackage[latin1]{inputenc}
\usepackage{amsfonts,amssymb,amsmath, epsfig}
\usepackage{color,graphicx,graphics}
\usepackage{amsthm}
\usepackage{amsmath,amstext,amssymb,amsfonts, amscd}
\usepackage[lofdepth,lotdepth]{subfig}
\usepackage{xcolor}
\usepackage{hyperref}          

\def\Box{\leavevmode\vbox{\hrule
     \hbox{\vrule\kern4pt\vbox{\kern4pt}%
           \vrule}\hrule}}
\def\blackbox{\leavevmode\vrule height 5pt width 4pt depth 0pt\relax}
\def\endproof{\null\hfill {$\blackbox$}\bigskip}

\newcounter{appendix}
\def\appendix{\advance\c@appendix by 1
   \def\thesection{\Alph{section}}
   \ifnum\c@appendix=1 \setcounter{section}{-1} \fi
   \@startsection {section}{1}{\z@}{-3.5ex plus -1ex minus 
   -.2ex}{2.3ex plus .2ex}{\Large\bf}}

\def\paragraph#1{{\bf #1\ }}

\newcommand{\rot}{\Theta}

\newtheorem{lemma}{Lemma}[section]  

\newtheorem{theorem}[lemma]{Theorem}

\newtheorem{proposition}[lemma]{Proposition}

\newtheorem{remark}{Remark}[section]

\title{Body-attitude coordination in arbitrary dimension} 
\author{P. Degond$^{(1)}$, A. Diez$^{(2)}$, A. Frouvelle$^{(3)}$} 
\date{} 
\begin{document}

\maketitle


\begin{center}
$^{(1)}$ Institut de Math\'ematiques de Toulouse ; UMR5219 \\
Universit\'e de Toulouse ; CNRS \\
UPS, F-31062 Toulouse Cedex 9, France\\
email: pierre.degond@math.univ-toulouse.fr

\bigskip

$^{(2)}$ Department of Mathematics, Imperial College London, South Kensington Campus \\
London, SW7 2AZ, United Kingdom\\
email: antoine.diez18@imperial.ac.uk

\bigskip

$^{(3)}$ CEREMADE, CNRS, Universit\'e Paris-Dauphine\\
Universit\'e PSL, 75016 Paris, France \\
email: frouvelle@ceremade.dauphine.fr

and

CNRS, Universit\'e de Poitiers, UMR 7348\\
Laboratoire de Math\'ematiques et Applications (LMA), 86000 Poitiers, France\\
email: amic.frouvelle@math.univ-poitiers.fr

\end{center}

\vspace{0.5 cm}
\begin{abstract}
We consider a system of self-propelled agents interacting through body attitude coordination in arbitrary dimension $n \geq 3$. We derive the formal kinetic and hydrodynamic limits for this model. Previous literature was restricted to dimension $n=3$ only and relied on parametrizations of the rotation group that are only valid in dimension $3$. To extend the result to arbitrary dimensions $n \geq 3$, we develop a different strategy based on Lie group representations and the Weyl integration formula. These results open the way to the study of the resulting hydrodynamic model (the ``Self-Organized Hydrodynamics for Body orientation (SOHB)'') in arbitrary dimensions.  
\end{abstract}

\medskip
\noindent
{\bf Acknowledgements:} PD holds a visiting professor association with the Department of
Mathematics, Imperial College London, UK. The work of AD is supported by an EPSRC-Roth scholarship cofounded by the Engineering and Physical Sciences Research Council and the Department of Mathematics at Imperial College London. AF acknowledges support from the Project EFI ANR-17-CE40-0030 of the French National Research Agency.


\vskip 0.4cm

\setcounter{equation}{0}
\section{Introduction}
\label{sec_intro}

Collective dynamics is an ubiquitous phenomenon in the living world at all scales (see e.g. locust swarms \cite{bazazi2008collective}, wildebeest herds \cite{shellard2020rules}, fish schools \cite{lopez2012behavioural}, human crowds \cite{warren2018collective},  bacterial colonies  \cite{be2019statistical}, embryogenic cell migration  \cite{giniunaite2020modelling}, semen \cite{creppy2016symmetry} or collective motion of subcellular structures \cite{saito2017understanding}). It produces large-scale structures (flocks, herds, coordinated motion) which extend far beyond the individual scale. A lot of effort has been put in understanding how local interactions between individuals can produce coordination over such large scales. Most approaches postulate some interaction rules at the individual level and assess them against experimental observations via large-scale computer simulations. 

One of the first and most widely used such models is the so-called three-zone model proposed by I. Aoki \cite{aoki1982simulation}, where individuals are assumed to attract each other at large distances, repel at short distance and align in the middle range. A simplified version of this model where individuals move at constant speed and are only subject to alignment has then been brought forward by T. Vicsek and his collaborators \cite{vicsek1995novel}. Independently, F. Cucker and S. Smale have developed a similar alignment model but with particles with variable speeds \cite{cucker2007emergent}. All these models have stimulated an intense literature which is impossible to cite exhaustively (see e.g. \cite{cao2020asymptotic} for the three-zone model, \cite{chate2008collective, costanzo2018spontaneous, degond2013macroscopic, degond2015phase, figalli2018global, frouvelle2012dynamics, gamba2016global, griette2019kinetic, jiang2016hydrodynamic, toner1998flocks, zhang2017local} for the Vicsek model, and \cite{aceves2019hydrodynamic, barbaro2016phase, barbaro2012phase, carrillo2010asymptotic, ha2009simple, ha2008particle, haskovec2021simple} for the Cucker-Smale model). Variants or combinations of these different models can be found in \cite{bertin2006boltzmann, bertin2009hydrodynamic, bertozzi2015ring, motsch2011new}. Connections between the Cucker-Smale and the Vicsek models via singular limits has been established in \cite{barbaro2012phase, bostan2013asymptotic, bostan2017reduced}. 

In these models, agents control their motion via the velocity or velocity director. A question is whether new behavior could be observed if more complex control variables are introduced. Recently, a new model in which the agents control their motion by their body attitude has been proposed in dimension $n=3$ \cite{Degond_eal_JNLS20, Degond_etal_arxiv21, degond2017new, Degond_etal_proc19, Degond_etal_MMS18, frouvelle2020body}. In this model, a reference frame describing the body attitude is attached to each agent. The agents move at constant speed in the direction of the first basis vector of the frame. The frame is updated to adjust to the ``average'' body attitude of the neighbors up to some noise. There are similarities with the Vicsek model except in the alignment process, which operates on the full body frame rather than on the single velocity vector. Consequently, heterogeneities in the body frame distribution affect motion and alter the dynamics compared with the Vicsek model \cite{Degond_etal_arxiv21}. So far, previous works were restricted to dimension $n=3$. The present paper is the first investigation of this body attitude coordination model in arbitrary dimensions~$n \geq 3$. 

Traditionally, there are three levels of modelling for systems of interacting agents. The finer level of description is the ``particle'' level, by which each agent is followed in the course of time by means of ordinary differential equations or stochastic processes \cite{aoki1982simulation, cao2020asymptotic, chate2008collective, costanzo2018spontaneous, cucker2007emergent, haskovec2021simple, lopez2012behavioural, motsch2011new, vicsek1995novel}. This is an appealing approach as the behavioral rules can be directed encoded in the equations. Due to the large number of particles, these models are not amenable to theoretical qualitative analyses. Thus, computer experiments are needed but computational efficiency drops dramatically when the number of particles becomes large. 

Then, a coarser approach consists of looking at the system in some statistical sense. Specifically, one monitors the evolution of the probability distribution of a representative particle. This leads to the so-called ``kinetic'' models, which are partial differential equations posed on high-dimensional space consisting of all the positional and control variables of the particles. This approach is mostly developed for theoretical purposes 
\cite{barbaro2016phase, bertin2006boltzmann, carrillo2010asymptotic, Degond_eal_JNLS20, degond2013macroscopic, figalli2018global, frouvelle2012dynamics, gamba2016global} although some numerical simulations have be attempted in low dimensional cases \cite{griette2019kinetic}. 

Finally the coarsest approach consists of describing the system by average or ``hydrodynamic'' quantities, such as the mean density, mean velocity, mean orientation, etc. In these hydrodynamic models, the agents control variables have been replaced by their local means which only depend on position. This results in considerable information savings and make the models amenable to both intensive numerical simulations and theoretical qualitative analysis \cite{bertin2009hydrodynamic, bertozzi2015ring, Degond_etal_arxiv21, toner1998flocks}.

By going from particle to kinetic then hydrodynamic models, microscopic information is gradually lost. Therefore, the validity of the kinetic and hydrodyanmic models compared with the particle ones is a key question. In model cases, it can be proved that kinetic models are a valid approximation of particle models, when the number of particles tends to infinity and a property named ``propagation of chaos'' is satisfied (see e.g. \cite{bolley2012mean, briant2020cauchy} for the Vicsek model, \cite{ha2009simple} for the Cucker-Smale model and \cite{diez2020propagation} for a general framework encompassing the body orientation model). Then, hydrodynamic models can be formally and rigorously derived from kinetic ones when the typical interaction time between the particles is small compared to the duration of the experiment (see \cite{degond2013macroscopic, degond2015phase, degond2008continuum, frouvelle2012continuum, jiang2016hydrodynamic, zhang2017local} for the Vicsek model and \cite{aceves2019hydrodynamic, barbaro2012phase} for the Cucker-Smale model). References \cite{cercignani2013mathematical, Degond_etal_proc19, giniunaite2020modelling, ha2008particle} give an overview of the whole process from particles to hydrodynamic equations. More phenomenological approaches can be found in \cite{bertin2006boltzmann, bertin2009hydrodynamic,  toner1998flocks}. 

For the body orientation model, the passage from kinetic to hydrodynamic equations has been realized in \cite{Degond_eal_JNLS20, Degond_etal_arxiv21, degond2017new, Degond_etal_proc19, Degond_etal_MMS18, frouvelle2020body} \textbf{in dimension $n=3$ only}. Specifically, in \cite{degond2017new}, the particle dynamics was described by a stochastic differential equation leading to a kinetic equation of Fokker-Planck type. In \cite{Degond_etal_MMS18} the representation of three-dimensional rotations by unit quaternions was used. In \cite{Degond_etal_proc19} instead, the particle dynamics was a jump process described by a piecewise deterministic Markov process (PDMP) resulting in a Boltzmann-BGK type kinetic equation. In \cite{Degond_eal_JNLS20, frouvelle2020body} this kinetic equation was studied from the viewpoint of phase transitions. Numerical comparisons of the PDMP with the hydrodynamic models were performed in \cite{Degond_etal_arxiv21}. In the present paper, we aim to derive the macroscopic equations corresponding to the same PDMP and Boltzmann-BGK kinetic equation as in \cite{Degond_etal_arxiv21} but in arbitrary dimensions $n \geq 3$. The passage from dimension $3$ to arbitrary dimensions is not a simple exercise. The reason is that all previous results obtained in dimension $3$ used a parametrization of the rotation group by means of either the Rodrigues formula or (equivalently) the unit quaternions. This parametrization is specific to dimension $n=3$ and does not extend to higher dimensions. Thus, it is necessary to use a completely different approach. In this paper, our main tools will be representation theory \cite{Fulton_Harris} and the so-called Weyl integration formula \cite{Simon}.

The search for a model in dimension $n >3$ may appear as a pointless endeavour as collectively moving agents (birds, fish) live in a three-dimensional world. In spite of this, there are advantages in deriving the model in arbitrary dimensions. First, the use of concepts specific to dimension 3 (such as cross product, curl, etc.) may hide the underlying structures of the model. By contrast, writing the model in arbitrary dimension may help decipher such structures (see \cite{charbonneau2013dimensional} for an elaboration of this argument). Finally, these models could potentially apply to more abstract entities, such as data. Collective dynamics models are increasingly used in optimization and data science \cite{clarte2019collective, pinnau2017consensus}. The present body attitude coordination model or a variant of it could be used as a building block for treating data structured in rotation matrices. 

Previous models involving agents' body attitudes have been proposed to model bird flocks \cite{hildenbrandt2010self}. However, there, the agents' interactions are similar to \cite{aoki1982simulation} and the body attitude itself is used to incorporate elements of flight physics in an individual agent's trajectory. The interplay between geometry and collective dynamics has recently appeared in the literature. In \cite{ahn2021emergent, fetecau2021well, fetecau2019self, sarlette2009consensus, sepulchre2010consensus}, collective dynamics models are considered where the particles are located on generic manifolds. Restrictions to specific manifolds are considered. For instance, the Lie group of unitary matrices is considered in \cite{golse2019mean}, more general Lie groups in \cite{ha2017emergent, sarlette2010coordinated} and the rotation group in 3 dimensions in \cite{fetecau2021emergent, sarlette2009autonomous}. However, the target is fairly different from ours. In these cases, the particles move in this manifold and are subjected to generalizations of the Cucker-Smale alignment dynamics (or extensions of the Kuramoto synchronization dynamics). In the present work, particles move in classical linear space but their control variables are changed into objects in a more complex geometric structure.  

The outline of this paper is as follows. Section \ref{sec_model} is devoted to the description of the modelling context, namely, the underlying particle model, the derivation of the kinetic model and its hydrodynamic scaling. The main result, namely, the formal convergence of the kinetic model towards the hydrodynamic model, is stated and discussed in Section \ref{sec_result}. The proof of the main result occupies Section \ref{sec_proof_main_thm}. A conclusion and perspectives are drawn in Section \ref{sec_conclu}. The proofs of several lemmas relying on representation theory are given in appendices. A summary of the results coming from representation theory used in these proofs is given at the beginning of the appendix.

\setcounter{equation}{0}
\section{Modelling context}
\label{sec_model}

\subsection{The particle system}
\label{subsec_particle}

Our starting point is the following particle system, which is a generalization of the three-dimensional model presented in \cite{Degond_etal_arxiv21, Degond_etal_proc19} to arbitrary dimensions $n \geq 3$. We assume the system composed of $N$ particles occupying spatial locations $X_k(t) \in {\mathbb R}^n$, $k \in \{1, \ldots, N\}$ in the course of time $t \geq 0$. Each particle is endowed with a moving direct orthonormal frame $(\omega^k_1(t), \ldots, \omega^k_n(t))$ which encodes its body attitude, or local body frame. Equivalently, letting $(e_1, \ldots, e_n)$ be the canonical basis of ${\mathbb R}^n$, we denote by $A_k(t)$ the unique rotation in $\mathrm{SO}_n {\mathbb R}$ which maps $(e_1, \ldots, e_n)$ to $(\omega^k_1(t), \ldots, \omega^k_n(t))$. The dynamics of the particles is a jump process. Between two jumps, the particles move in straight line at uniform speed $c_0$ (supposed identical for all the particles) in the direction of the first basis vector $\omega^k_1(t) = A_k(t) e_1$ of the local body frame (also referred to as the self-propulsion direction), while the body frame $A_k(t)$ remains unchanged. 

Jump times are exponentially distributed: they form an increasing random sequence $T_k^1, \, T_k^2, \ldots$ such that $T_k^{j+1} - T_k^j$ follows a Poisson law with constant parameter $\nu>0$. At jump times $T_k^j$, $X_k$ is continuous but $A_k$ experiences a jump between $A_k(T_k^j-0)$ and $A_k(T_k^j+0)$. To define how $A_k$ jumps, we first need to define the neighbor's average frame. Omitting the time variable for simplicity, we first introduce the following average
\begin{equation}
J_k = \frac{1}{N} \sum_{\ell=1}^N K(X_k-X_\ell) \, A_\ell,  
\label{eq:defJk}
\end{equation}
where the function $K$ (the sensing function): ${\mathbb R}^n \to [0,\infty)$ is given once for all. For simplicity, we will assume that $K$ is a radial function: 
\begin{equation} 
K(x) = \frac{1}{R^n} \tilde K(\frac{|x|}{R}), 
\label{eq:sensing}
\end{equation}
(where $|x|$ is the euclidean norm of $x$) for a suitable function $\tilde K$: $[0,\infty) \to [0,\infty)$ and a spatial scale parameter $R>0$ referred to as the sensing radius. A typical example of a function $\tilde K$ is the indicator of the interval $[0,1]$. Since $J_k$ is not a rotation matrix in general, we project it onto $\mathrm{SO}_n {\mathbb R}$ using the formula
\begin{equation}
\label{projection0}
\rot_k:= \textrm{arg  max}_{A \in \mathrm{SO}_n {\mathbb R}} \, A \cdot J_k , 
\end{equation}
where, for two matrices $A$, $B$ in ${\mathcal M}_n({\mathbb R})$, the space of $n \times n$ matrices with real entries,  
\begin{equation}
A \cdot B = \mathrm{Tr} \{ A^T B \} = \sum_{i,j=1}^n A_{ij} B_{ij}, 
\label{eq:innerprod}
\end{equation}
is the standard inner product on ${\mathcal M}_n({\mathbb R})$, Tr stands for the trace and the exponent $T$ for the matrix transpose. Eq. \eqref{projection0} defines $\rot_k$ as the element $A$ that maximizes $A \cdot J_k$ over $\mathrm{SO}_n {\mathbb R}$. Indeed, this element is unique if and only if $J_k$ is not singular (i.e. $\det J_k \not = 0$ where $\det$ stands for the determinant), which we will suppose from now on. The set of singular matrices being of zero-measure, it is indeed legitimate to expect that such a singularity will not happen except for a negligible set of initial conditions. Moreover, if $\det J_k > 0$, $\rot_k$ coincides with the unique rotation provided by the polar decomposition of $J_k$. The rotation $\rot_k$ is what we define as the neighbor's average frame. Now, we can define how $A_k(T_k^j-0)$ jumps to $A_k(T_k^j+0)$. We first define $J_k(T_k^j-0)$ by \eqref{eq:defJk} with $X_k$ and $X_\ell$ taking the values $X_k(T_k^j-0)$ and $X_\ell(T_k^j-0)$ (note that in general, for $\ell \not = k$, $X_\ell$ will be continuous at $T_k^j$). We deduce $\rot_k (T_k^j-0)$ from \eqref{projection0}. Then $A_k(T_k^j+0)$ is sampled according to the probability distribution $M_{\rot_k (T_k^j-0)}$ on $\mathrm{SO}_n {\mathbb R}$, where, for any $\rot \in \mathrm{SO}_n {\mathbb R}$, the function $M_{\rot}$: $\mathrm{SO}_n {\mathbb R} \to {\mathbb R}$ is the von-Mises probability density: 
\begin{equation}
M_{\rot} (A) = \frac{1}{Z} \exp (\kappa \, \rot \cdot A), \quad Z = \int_{\mathrm{SO}_n {\mathbb R}} \exp \big( \kappa \, \mbox{Tr} A \big)  \, dA.
\label{eq:VM}
\end{equation}
Here, $dA$ stands for the normalized Haar measure on $\mathrm{SO}_n {\mathbb R}$ (we recall that, for $\mathrm{SO}_n {\mathbb R}$ as for any compact group, the Haar measure is invariant by both left and right translations, by conjugation and by group inversion). The parameter $\kappa$ is the concentration parameter of the von Mises and plays the role of the inverse of the temperature. It is supposed given and constant. The rotation $\rot$ is called the orientation parameter of the von Mises. 

In summary, the particle dynamics is a Piecewise Deterministic Markov Process (PDMP) described as follows. 
\begin{itemize}
\item Between jump times i.e. for all $t\in[T_k^j,T_k^{j+1})$, we have 
\begin{equation}
\label{deterministicpart}
X_k(t) = X_k(T_k^j) + c_0 \, (t-T_k^j) \, A_k(t) \, e_1,  \quad A_k(t)=A_k(T_k^j+0).
\end{equation}
\item At $T_k^j$, $X_k$ is continuous and $A_k$ jumps from $A_k(T_k^j-0)$ to $A_k(T_k^j+0)$,  where $A_k(T_k^j+0)$ is drawn from a von Mises distribution, i.e.  
\begin{equation}
\label{jumppart}
A_k(T_k^j+0) \sim M_{\rot_k (T_k^j-0)}.
\end{equation}
\end{itemize}

\begin{figure}[ht!]
\centering
\subfloat[Particle jump process]{\includegraphics[trim={3.5cm 21.5cm 10.5cm 4.5cm},clip,width=8.5cm]{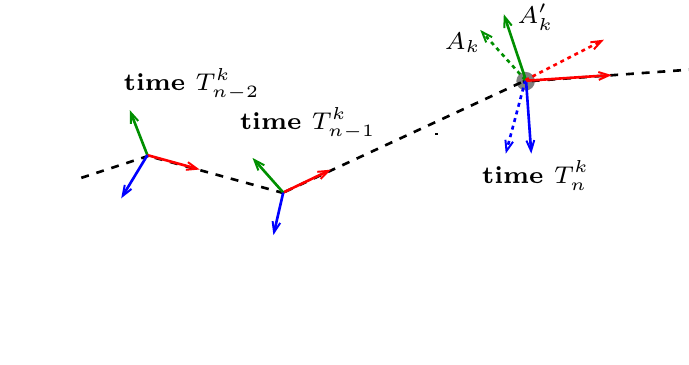}\label{fig:jump1}}
\subfloat[Detail of a jump]{\includegraphics[trim={3.5cm 22.5cm 12.5cm 3.5cm},clip,width= 9cm]{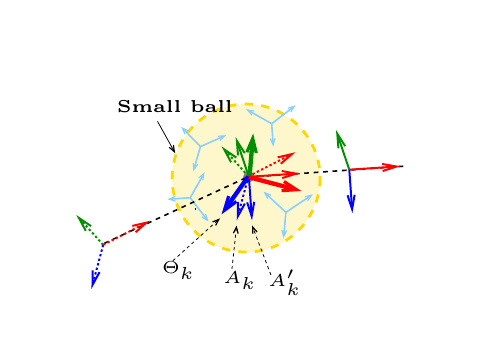}\label{fig:jump2}}
\caption{(a) Graphical representation of the particle jump process: between the jump times $T_{n}^k$, the body frame (represented by red, green and blue arrows) remains constant and the particle trajectory (the dashed black line) is a straight line directed by the first basis vector of the body frame (in red).  At jump times, the body frame experiences a sudden change while the trajectory line is broken and a different straight line begins. (b)~Detail of a jump: the body frame jumps from its value $A_k$ (depicted by thin dashed arrows) to a new value $A_k'$ (depicted by thin solid arrows) sampled from a von Mises distribution whose orientation parameter is the frame $\rot_k$ (depicted by thick arrows). The orientation parameter is obtained by averaging the body frame of the neighboring particles (in blue), i.e. those contained in a small ball (in yellow) centered at the jump location. }
\label{fig:jump}
\end{figure}

We refer to Fig. \ref{fig:jump} for a graphical depiction of the particle dynamics. Ref. \cite{Degond_etal_arxiv21} provides a link to numerical simulations of this model in dimension $n=3$ made on graphics processor units (GPU) using the SiSyPHE package \cite{diez2021sisyphe}.

\begin{remark} 
The normalization constant (aka the partition function) $Z$ in \eqref{eq:VM} should be defined by 
$$ Z = \int_{\mathrm{SO}_n {\mathbb R}} \exp \big( \kappa \rot \cdot A \big)  \, dA. $$
However, we can use the change of variables $A' = \rot^T A$ and the left invariance of the Haar measure to derive the expression of $Z$ in \eqref{eq:VM}, which incidentally shows that the normalization constant of $M_{\rot}$ does not depend on $\rot$. 
\end{remark}

\subsection{The kinetic equation}
\label{subsec_kinetic}

A kinetic equation can formally be derived in the limit $N \to \infty$. To justify this, let us temporarily modify the particle model by changing \eqref{jumppart} into 
\begin{equation}
\label{jumppart_mod}
A_k(T_k^j+0) \sim M_{J_k (T_k^j-0)}.
\end{equation}
Indeed, we note from \eqref{eq:VM} that the von Mises distribution is well-defined for any matrix $\rot$ and not just for $\rot \in \mathrm{SO}_n {\mathbb R}$. By this modification, we skip the normalization step and we do not need to assume that $J_k$ is non-singular. In this case, it has been shown in \cite{diez2020propagation} that, under some hypotheses on the initial conditions, the particle system is well-defined for all time and that the random measure 
$$ \mu^N(x,A,t) = \frac{1}{N} \sum_{k=1}^N \delta_{(X_k(t), A_k(t))}(x,A), $$
where $\delta_{(X_k(t), A_k(t))}(x,A)$ stands for the Dirac delta in ${\mathbb R}^N \times \mathrm{SO}_n {\mathbb R}$ located at $(X_k(t), A_k(t))$, converges as $N \to \infty$ in weak sense to a deterministic absolutely continuous measure $f(x,A,t)$ which satisfies a kinetic equation. 

Although the extension of this result to the present case is still an open (and likely difficult) problem, we will assume that this result is still true. In this case, the equation satisfied by $f$ is given by the following kinetic model
\begin{equation}
\partial_t f + (c_0 A e_1 \cdot \nabla_x) f = \nu \,  \big( \tilde \rho_f M_{\tilde \rot_f} - f \big) . 
\label{eq:BGK_unsc}
\end{equation} 
We remind that $A e_1$ is the first column vector of the matrix $A$.  For a given distribution function $f=f(x,A)$, the function $\tilde \rho_f$: ${\mathbb R}^n \to [0,\infty)$ is the non-local density 
$$ \tilde \rho_f (x) = \int_{{\mathbb R}^n \times \mathrm{SO}_n {\mathbb R}} K(x-y) \, f(y, A) \, dy \, dA. $$
The function  $\tilde \rot_f$: ${\mathbb R}^n \to \mathrm{SO}_n {\mathbb R}$ is defined from \eqref{projection0} where $J_k$ is replaced by $\tilde J_f$ given by 
$$ \tilde J_f(x) =  \int_{{\mathbb R}^n \times \mathrm{SO}_n {\mathbb R}} K(x-y) \, f(y, A) \, A \, dy \, dA. $$
Again, if $\tilde J_f$ has non-zero determinant, which we will suppose throughout this paper, $\tilde \rot_f$ is uniquely defined and if $\tilde J_f$ has positive determinant, $\tilde \rot_f$ coincides with the orthogonal matrix involved in the polar decomposition of $\tilde J_f$. Of course, if $f$ also depends on $t$, $\tilde \rho_f$ and $\tilde \rot_f$ also do.  Note that, in $A e_1 \cdot \nabla_x$, the dot simply means the inner product of two vectors of ${\mathbb R}^n$, namely $A e_1$ and the differential symbol $\nabla_x$. The same notation will be used for inner products of vectors and matrices as the context will always be clear.

\begin{remark}
If we had used the modification \eqref{jumppart_mod}, $\tilde \rot_f$ in \eqref{eq:BGK_unsc} would have been replaced by $\tilde J_f$. In that case, the result of \cite{diez2020propagation} would apply and we would have the rigorous convergence of $\mu^N$ to $f$ as $N \to \infty$. However, the resulting macroscopic model would be more complicated. This is why we choose to work with the normalized model in spite of the absence of rigorous justification of the kinetic model. 
\end{remark}

\subsection{Non-dimensionalization and scaling}
\label{subsec_nondim}

We now define a time scale $t_0$ and the associated space scale $x_0 = c_0 t_0$. We introduce two dimensionless parameters 
$$  \bar \nu = \nu t_0, \qquad \bar R = \frac{R}{x_0}, $$
and define dimensionless variables $x' = x/x_0$, $t' = t/t_0$, and functions $f'(x',A,t') = x_0^n f(x,A,t)$, $\tilde \rho_{f'}' = x_0^n \tilde \rho_f(x,t)$,  $\tilde J_{f'}' = x_0^n \tilde J_f(x,t)$. In these new variables, Eq. \eqref{eq:BGK_unsc} is written (dropping the primes for simplicity):
\begin{equation}
\partial_t f + (A e_1 \cdot \nabla_x) f = \bar \nu \,  \big( \tilde \rho_f M_{\tilde \rot_f} - f \big) . 
\label{eq:BGK_sc}
\end{equation} 
with, using \eqref{eq:sensing}
\begin{eqnarray*}
\tilde \rho_f (x) &=& \int_{{\mathbb R}^n \times \mathrm{SO}_n {\mathbb R}} \frac{1}{\bar R^n} \tilde K \big(\frac{|x-y|}{\bar R} \Big) \, f(y, A) \, dy \, dA, \\
\tilde J_f(x) &=&   \int_{{\mathbb R}^n \times \mathrm{SO}_n {\mathbb R}} \frac{1}{\bar R^n} \tilde K \big(\frac{|x-y|}{\bar R} \Big) \, f(y, A) \, A \, dy \, dA.
\end{eqnarray*}

We now make the following hydrodynamic scaling assumptions:
$$ \frac{1}{\bar \nu} = \bar R =  \varepsilon \ll 1. $$
The parameter $\varepsilon >0$ is at the same time a dimensionless measure of the relaxation speed of $f$ towards the local equilibrium $\rho_f M_{\rot_f}$ (analog of the Knusdsen number in rarefied gases) and the typical scale of the interaction radius. An easy Taylor expansion and the fact that the sensing kernel is rotationally symmetric show that 
$$ \tilde \rho_f  = \rho_f + {\mathcal O}(\varepsilon^2), \quad \tilde J_f  = J_f + {\mathcal O}(\varepsilon^2), \quad \tilde \rot_f = \rot_f + {\mathcal O}(\varepsilon^2),$$
with 
\begin{equation}
\rho_f (x) = \int_{\mathrm{SO}_n {\mathbb R}} \, f(x, A) \, dA, \qquad 
J_f(x) =  \int_{\mathrm{SO}_n {\mathbb R}}  \, f(x, A) \, A \, dA,
\label{projection}
\end{equation}
and $\rot_f(x)$ deduced from $J_f(x)$ by \eqref{projection0} (replacing $J_k$ by $J_f(x)$). From \eqref{eq:BGK_sc}, we get 
$$
\partial_t f + (A e_1 \cdot \nabla_x) f = \frac{1}{\varepsilon} \,  \big( \rho_f M_{\rot_f} - f \big) + {\mathcal O}(\varepsilon),  
$$ 
and we drop the ${\mathcal O}(\varepsilon)$ remainder as it has no contribution to the result we are aiming at. 

Denoting the unknown by $f^\varepsilon (x,A,t)$ to highlight its dependence on $\varepsilon$, we are finally led to the following perturbation problem
\begin{equation}
\partial_t f^\varepsilon + (A e_1 \cdot \nabla_x) f^\varepsilon = \frac{1}{\varepsilon} \big( \rho_{f^\varepsilon} M_{\rot_{f^\varepsilon}} - f^\varepsilon \big) , 
\label{eq:BGK}
\end{equation} 
with $\rho_f$ and $J_f$ given by \eqref{projection}. In this paper, we address the problem of deriving the formal limit $\varepsilon \to 0$ of model \eqref{eq:BGK}

\setcounter{equation}{0}
\section{Main result and discussion}
\label{sec_result}

We first introduce some notations. For two vectors $X = (X_i)_{i=1, \ldots, n}$ and $Y = (Y_i)_{i=1, \ldots, n}$, $X \wedge Y$ and $\nabla_x \wedge X$ denote the antisymmetric matrices:
$$ (X \wedge Y)_{ij} = X_i Y_j - X_j Y_i, \quad (\nabla_x \wedge X)_{ij} = \partial_{x_i} X_j - \partial_{x_j} X_i, $$
(note that $X \wedge Y$ is the exterior product of $X$ and $Y$ and, if $X$ is identified with a $1$-form through the euclidean structure of ${\mathbb R}^n$, $\nabla_x \wedge X$ is the exterior derivative of $X$). 

For $k \in {\mathbb Z}$ and $p \in {\mathbb N} \setminus \{0\}$, we define the functions $C_{2p}^{(k)}$ and $C_{2p+1}^{(k)}$: ${\mathbb R}^p \to {\mathbb R}$ by
\begin{eqnarray}
C_{2p}^{(k)} (\theta_1, \ldots, \theta_p) &=& 2(\cos  k \theta_1 + \ldots + \cos  k \theta_p), \label{eq:Ck2p} \\
C_{2p+1}^{(k)} (\theta_1, \ldots, \theta_p) &=& 2(\cos  k \theta_1 + \ldots + \cos  k \theta_p) + 1.  \label{eq:Ck2p+1}
\end{eqnarray}
We also define $u_{2p}$ and $u_{2p+1}$: ${\mathbb R}^p \to {\mathbb R}$ by 
\begin{eqnarray}
\hspace{-0.5cm} u_{2p} (\theta_1, \ldots, \theta_p) &=& \frac{2^{(p-1)^2}}{p!} \prod_{1 \leq j<k \leq p} \big( \cos \theta_j - \cos \theta_k \big)^2, 
\label{eq:u2p} \\
\hspace{-0.5cm} u_{2p+1} (\theta_1, \ldots, \theta_p) &=& \frac{2^{p(p-1)}}{p!} \prod_{1 \leq j<k \leq p} \big( \cos \theta_j - \cos \theta_k \big)^2 \, \prod_{j=1}^p \big( 1 - \cos \theta_j \big). 
\label{eq:u2p+1}
\end{eqnarray}

The goal of this paper is to prove the formal theorem: 

\begin{theorem}
Let $n \in {\mathbb N}$, $n\geq 3$. We assume that \eqref{eq:BGK} has a smooth solution $f^\varepsilon$ which converges as $\varepsilon \to 0$ as smoothly as needed towards a smooth function $f$. Then, 
\begin{equation}
f = \rho M_{\rot}, 
\label{eq:cvgce}
\end{equation} 
where $\rho = \rho(x,t)$ and $\rot = \rot(x,t)$ are functions from ${\mathbb R}^n \times (0,\infty)$ to $(0,\infty)$ and $\mathrm{SO}_n {\mathbb R}$ respectively which are solutions of the following system: 
\begin{eqnarray}
&& \hspace{-1cm} 
\partial_t \rho + \nabla_x \cdot (c_1 \rho \Omega_1) = 0,  \label{eq:mass_2} \\
&& \hspace{-1cm} 
\rho \big( \partial_t + c_2 \, \Omega_1 \cdot \nabla_x \big) \rot = {\mathbb W} \rot,  
\label{eq:orient_3} 
\end{eqnarray}
where 
\begin{eqnarray}
{\mathbb W} &=& F \wedge \Omega_1 - c_4 \, \rho \, \nabla_x \wedge \Omega_1,  \label{eq:expressW} \\
F &=& - c_3 \, \nabla_x \rho - c_4 \, \rho \, r,  \label{eq:expressF} \\
r &=& \sum_{k=1}^n (\nabla_x \cdot \Omega_k) \Omega_k, \label{eq:expressr}
\end{eqnarray} 
with $ \Omega_k = \rot e_k$, $k = 1, \ldots, n$. The constants $c_i$, $i=1, \ldots, 4$ are expressed as follows. Let $p \in {\mathbb N}$ such that $n=2p$ or $n=2p+1$. We have 
\begin{eqnarray}
c_1 &=& \frac{1}{n} \frac{\displaystyle \int_{[0,2\pi]^p} C_n^{(1)} \, \exp \big( \kappa C_n^{(1)} \big) \, u_n \, d \tilde \theta_p}{\displaystyle \int_{[0,2\pi]^p} \exp \big( \kappa C_n^{(1)} \big) \, u_n \,  d \tilde \theta_p}, 
\label{eq:express_c1_WIF} \\
c_2 &=& \frac{\displaystyle \int_{[0,2\pi]^p} \big( 2 C_n^{(3)} - n C_n^{(1)} \, C_n^{(2)} + (n^2-2) C_n^{(1)} \big)  \, \exp \big( \kappa C_n^{(1)} \big) \, u_n \, d \tilde \theta_p}{\displaystyle n (n-2)(n+2) \, \int_{[0,2\pi]^p} \big( 1- \frac{1}{n} C_n^{(2)} \big)  \, \exp \big( \kappa C_n^{(1)} \big) \, u_n \, d \tilde \theta_p}, 
\label{eq:express_c2_WIF} \\
c_3 &=& \frac{1}{2 \kappa}, \label{eq:express_c3} \\
c_4 &=& \frac{\displaystyle \int_{[0,2\pi]^p} \big( C_n^{(3)} - \frac{2}{n} C_n^{(1)} \, C_n^{(2)} + C_n^{(1)} \big)  \, \exp \big( \kappa C_n^{(1)} \big) \, u_n \, d \tilde \theta_p}{\displaystyle 2(n-2)(n+2) \, \int_{[0,2\pi]^p} \big( 1- \frac{1}{n} C_n^{(2)} \big)  \, \exp \big( \kappa C_n^{(1)} \big) \, u_n \, d \tilde \theta_p},  
\label{eq:express_c4_WIF}
\end{eqnarray}
with the notation $d \tilde \theta_p = d \theta_1 \ldots d\theta_p$. Furthermore, The function $c_1$: $[0,\infty) \to {\mathbb R}$, $\kappa \mapsto c_1(\kappa)$ is non-decreasing and satisfies $c_1(0) = 0$ and $\lim_{\kappa \to \infty} c_1(\kappa) = 1$. Thus, it is a  bijection from $[0,\infty)$ to $[0,1)$.
\label{thm:main}
\end{theorem}

There is an alternate expression of \eqref{eq:orient_3} which requires the introduction of additional notations. Suppose $A$, $B$ and $C$ are three smooth vector fields ${\mathbb R}^n \to {\mathbb R}$. We define 
\begin{equation}
\delta(A,B,C) = \big( (A \cdot \nabla_x) B \big) \cdot C + \big( (B \cdot \nabla_x) C \big) \cdot A + \big( (C \cdot \nabla_x) A \big) \cdot B, 
\label{eq:def_delta}
\end{equation}
and 
\begin{equation}
\Delta_{ijk} = \delta(\Omega_i, \Omega_j, \Omega_k). 
\label{eq:def_delijk}
\end{equation}
It is easy to check that $\Delta_{ijk}$ is antisymmetric by permutations of the indices $(i,j,k)$. Then, we define the following antisymmetric matrix field:
\begin{equation}
{\mathbb A} = \sum_{k,\ell = 1}^n \Delta_{1k\ell} \, \Omega_k \otimes \Omega_\ell, 
\label{eq:defmathbbA}
\end{equation}
where $\otimes$ denotes the tensor product of two vectors. The matrix ${\mathbb A}$ is just the matrix with entries $\Delta_{1k\ell}$ in the basis $(\Omega_1, \ldots, \Omega_n)$. We note that 
\begin{equation}
{\mathbb A} \Omega_1 = \sum_{k,\ell = 1}^n \Delta_{1k\ell} \, \Omega_k \, (\Omega_\ell \cdot \Omega_1) = \sum_{k= 1}^n \Delta_{1k1} \, \Omega_k = 0, 
\label{eq:AOm1=0}
\end{equation}
by the antisymmetry of $\Delta_{ijk}$. Finally, we define
\begin{equation}
\tilde {\mathbb W} = F \wedge \Omega_1 + c_4 \rho  {\mathbb A}. 
\label{eq:deftilW}
\end{equation}
Then, we have the following proposition, whose proof can be found in Appendix \ref{sec_comp3D}. 

\begin{proposition}
Eq. \eqref{eq:orient_3} is equivalent to 
\begin{equation}
\rho \big( \partial_t + (c_2-c_4) \, \Omega_1 \cdot \nabla_x \big) \rot = \tilde {\mathbb W} \rot.  
\label{eq:orient_4} 
\end{equation}
\label{prop:alterrot}
\end{proposition}

Eq. \eqref{eq:orient_4} can be expanded into equations for the basis vectors $\Omega_j$. We have:
\begin{eqnarray}
&&\hspace{-1cm}
\rho \big( \partial_t + (c_2-c_4) \, \Omega_1 \cdot \nabla_x \big) \Omega_1 = -c_3 P_{\Omega_1^\bot} \nabla_x \rho - c_4 \rho \sum_{k \not = 1} (\nabla_x \cdot \Omega_k) \Omega_k , \label{eq:Om1} \\
&&\hspace{-1cm}
\rho \big( \partial_t + (c_2-c_4) \, \Omega_1 \cdot \nabla_x \big) \Omega_j = \big( c_3 (\Omega_j \cdot \nabla_x) \rho + c_4 \rho (\nabla_x \cdot \Omega_j) \big) \Omega_1 \nonumber \\
&&\hspace{4cm}
- c_4 \rho \sum_{k \not \in \{1,j\}} \delta(\Omega_1, \Omega_j, \Omega_k) \Omega_k, \qquad j = 2, \ldots, n, \label{eq:Omk}
\end{eqnarray}
where $P_{\Omega_1^\bot} = \mathrm{Id} - \Omega_1 \otimes \Omega_1$ is the orthogonal projection onto $\{\Omega_1\}^\bot$. 

Below and in Appendix \ref{sec_comp3D}, we show that, in dimension $n=3$, this system coincides with the so-called '`Self-Organized Hydrodynamics for Body-orientation (SOHB)'' derived earlier \cite{Degond_eal_JNLS20, degond2017new}. This model is a nonlinear nonconservative hyperbolic system \cite{Degond_etal_inprep} composed of two equations, a mass conservation (or continuity) equation \eqref{eq:mass_2} for the density $\rho$ and an evolution equation \eqref{eq:orient_4} for the mean body orientation $\rot$. 

The continuity equation \eqref{eq:mass_2} shows that the fluid velocity is $c_1 \Omega_1$, i.e. the fluid flows in the direction of the first basis vector of the mean body orientation matrix $\rot$, at a speed equal to $c_1$. As $c_1 \in [0,1]$, the fluid speed is slower than the particle speed. This is due to the dispersion of the particle velocities $A e_1$ around the mean velocity $\Omega_1$ and the fact that the norm of the average is less than the average of the norms. This speed reduction is all the greater than the noise intensity $\kappa^{-1}$ is greater, as the increase of $c_1$ with $\kappa$ shows ($c_1$ is an order parameter \cite{Degond_eal_JNLS20}). This type of continuity equation is classical in collective dynamics, and has been also found e.g. in the Vicsek model \cite{degond2008continuum}. 

The left-hand side of \eqref{eq:orient_3} is a convective derivative along the vector field $(c_2-c_4) \Omega_1$. As $c_2 -c_4 \not = c_1$, this convective derivative is not a material derivative, i.e. it is not the derivative of $\rot$ when a fluid element is followed in its motion. Rather, one needs to move at speed $c_2-c_4$ different from the fluid speed $c_1$ along the velocity direction $\Omega_1$ to follow the evolution of $\rot$. Along this motion, $\rot$ can be seen as a moving frame. So, its (convective) derivative, by classical rigid-body dynamics, is obtained by multiplying it on the left by an antisymmetric matrix, which is the matrix $\tilde {\mathbb W}$. 

This matrix has two components. The first one is $F \wedge \Omega_1$. It describes an infinitesimal rotation in the plane Span$\{F,\Omega_1\}$ which tends to align $\Omega_1$ with $F$. In the orthogonal supplement $($Span$\{F,\Omega_1\})^\bot$ of the frame, this component of $\tilde {\mathbb W}$ does not produce any motion. By \eqref{eq:expressF}, the vector $F$ itself has two components. The first one $-c_3 \nabla_x \rho$ tends to turn the fluid velocity away from large density regions. It has the same effect as the pressure gradient in classical fluid dynamics. The second component of $F$ is proportional to the vector $r$. Its effect is to deviate the fluid in the presence of spatial gradients of the mean body attitude, an effect which has no counterpart in classical fluid dynamics. Novel effects brought by this additional term have been investigated in 3D in \cite{Degond_etal_arxiv21}. 

The second component of $\tilde {\mathbb W}$ is proportional to ${\mathbb A}$. The important feature of ${\mathbb A}$ is \eqref{eq:AOm1=0} which says that the self-propulsion direction $\Omega_1$ is not modified by ${\mathbb A}$. Hence ${\mathbb A}$ describes an infinitesimal rotation of the frame about the self-propulsion direction $\Omega_1$. Thus, \eqref{eq:deftilW} can be seen as the decomposition of the infinitesimal rotation $\tilde W$ into a component that fixes $\Omega_1$ and is independent of the force $F$ (that proportional to ${\mathbb A}$) and its complement (proportional to $F \wedge \Omega_1$) which embodies the effect of $F$. Again, the frame rotation about~$\Omega_1$ has no counterpart in classical fluid dynamics. Note that this decomposition has been made possible thanks to the choice of the ``right'' convection velocity $c_2 - c_4$. Expression \eqref{eq:expressW} for ${\mathbb W}$ does not have such a simple interpretation, which indicates that the speed $c_2$ used in \eqref{eq:orient_3} is not the most natural choice for the convection speed of $\rot$. Also, note that a single coefficient $c_4$ controls the two effects of the gradients of body attitude (that proportional to $r \wedge \Omega_1$ and that proportional to ${\mathbb A}$), suggesting that these two effects have a common microscopic origin. 

\begin{figure}[ht!]
\centering
\subfloat[Action of $F \wedge \Omega_1$ \hspace{0.2cm} $\mbox{}$]{\includegraphics[trim={3.5cm 21.5cm 11.5cm 4.5cm},clip,width=8cm]{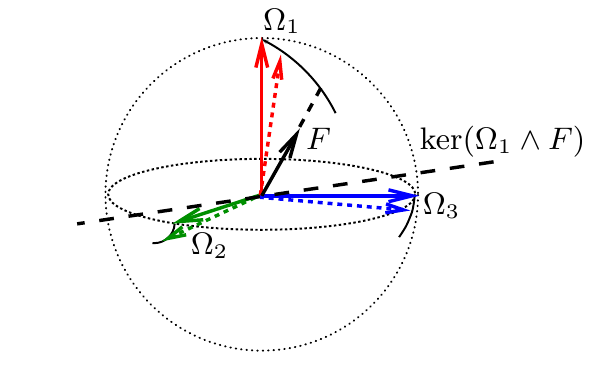}\label{fig:action_F}} 
\subfloat[Action of ${\mathbb A}$ \hspace{0.7cm} $\mbox{}$]{\includegraphics[trim={3.5cm 21.5cm 11.5cm 4.5cm},clip,width=8cm]{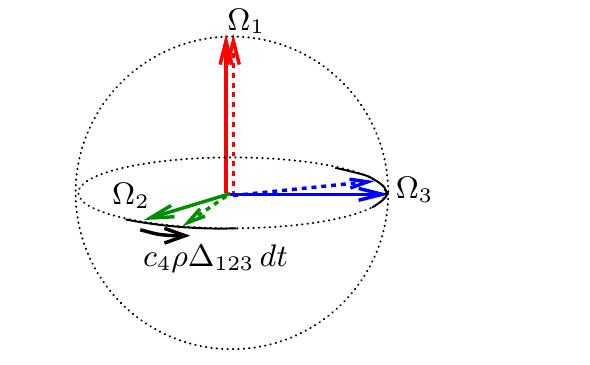}\label{fig:action_A}}
\caption{The effects of the two components of $\tilde {\mathbb W}$ on the motion of the frame, represented in dimension $n=3$ by red, green and blue arrows for $\Omega_1$ (the self-propulsion direction), $\Omega_2$ and $\Omega_3$ respectively. The solid arrows represent the frame at a time $t$ and the dashed arrows, at a later time $t + \delta t$ with $\delta t \ll 1$.  (a) Action of $F \wedge \Omega_1$ (assuming ${\mathbb A} = 0$): the force $F$ is the black arrow. The frame motion is depicted by the segment of circles joining the ends of the solid arrows and the dashed arrows. The frame rotates in the plane Span$\{F,\Omega_1\}$ to align $\Omega_1$ with $F$. The space $($Span$\{F,\Omega_1\})^\bot$ is the kernel of $F \wedge \Omega_1$ (the dashed line) so, this space is kept fixed in the frame motion. (b) Action of ${\mathbb A}$ (assuming $F=0$): the frame rotates about $\Omega_1$ which remains fixed. Vector $\Omega_2$ moves in the plane Span$\{\Omega_2,\Omega_3\}$ with angular speed $c_4 \rho \Delta_{123}$ (and similarly for $\Omega_3$). The displacement of the end of the arrow representing $\Omega_2$ is depicted by the black arrow. }
\label{fig:action}
\end{figure}

The effects of the two components of $\tilde {\mathbb W}$ are depicted in Fig. \ref{fig:action}.

Let us now focus on the dimension $n=3$ case and compare the results of the present paper with \cite{Degond_etal_arxiv21}. Indeed, the two papers feature the same individual-based model as a starting point, but for the restriction to dimension $n=3$ in \cite{Degond_etal_arxiv21}. The notations of \cite{Degond_etal_arxiv21} are slightly modified to fit with the notations used in the present paper. First, we introduce some notations specific to the dimension $n=3$. For a vector $w \in {\mathbb R}^3$, $[w]_\times$ denotes the antisymmetric matrix associated to the map ${\mathbb R}^3 \ni x \mapsto w \times x \in {\mathbb R}^3$. For two vector fields $F$ and $G$, we have $F \wedge G = - [F \times G]_\times$, where $F \times G$ stands for the vector product of $F$ and $G$. Introducing $\tilde \delta =: \Delta_{123}$, we have from \eqref{eq:defmathbbA}:
$$ {\mathbb A} = \tilde \delta (\Omega_2 \otimes \Omega_3 - \Omega_3 \otimes \Omega_2) = - \tilde \delta [\Omega_1]_\times. $$
Thus, \eqref{eq:deftilW} leads to
\begin{equation}
\tilde {\mathbb W} = - [\tilde w]_\times, \qquad \tilde w = F \times \Omega_1 + \tilde \delta \Omega_1. 
\label{eq:tilWn=3}
\end{equation}
Then, we can check that Eqs. \eqref{eq:mass_2}, \eqref{eq:orient_4} together with \eqref{eq:tilWn=3}, \eqref{eq:expressF} and \eqref{eq:expressr} are identical with Eqs. (20)-(22) of \cite{Degond_etal_arxiv21} (note that in \cite{Degond_etal_arxiv21} the constant $c_2$ is what we call $c_2 - c_4$). Thus, our result and that of \cite{Degond_etal_arxiv21} are compatible, provided we show that the expressions \eqref{eq:express_c1_WIF}-\eqref{eq:express_c4_WIF} of the constants $c_1$ to $c_4$ in dimension $n=3$ are the same as those found in \cite{Degond_etal_arxiv21}. This is indeed the case and the computation is given in Appendix \ref{sec_comp3D}.

\setcounter{equation}{0}
\section{Proof of Theorem \ref{thm:main}}
\label{sec_proof_main_thm}

\subsection{Limit $\varepsilon \to 0$}
\label{sec_epsto0}

We first need the following Lemma whose proof is given in Appendix \ref{sec:proof:int_A_VM_dA}. It uses the following notation:
$$ \langle f(A) \rangle_{g(A)} = \frac{\int_{\mathrm{SO}_n {\mathbb R}} f(A) g(A) \, dA}{\int_{\mathrm{SO}_n {\mathbb R}} g(A) \, dA}, $$
where $f$ and $g$ are two functions $\mathrm{SO}_n {\mathbb R} \to {\mathbb R}$, $g \geq 0$ and $g \not \equiv 0$.

\begin{lemma}
We have 
\begin{equation}
\int_{\mathrm{SO}_n {\mathbb R}}  A \, M_{\rot}(A) \, dA = c_1 \, \rot, \quad c_1 =  \big \langle \frac{\mathrm{Tr} A}{n} \big \rangle_{\exp (\kappa \mathrm{Tr} A )}. 
\label{eq:int_A_VM_dA}
\end{equation}
$c_1$ is a non-decreasing and satisfies $c_1(0) = 0$ and $\lim_{\kappa \to \infty} c_1(\kappa) = 1$. 
\label{lem:int_A_VM_dA}
\end{lemma}

\medskip
Next, for $f$: $\mathrm{SO}_n {\mathbb R} \to [0,\infty)$ we define the collision operator
$$ Q(f) = \rho_{f} M_{\rot_f} - f,  $$
The following lemma gives the equilibria of $Q$:

\begin{lemma}
For functions $f$: $\mathrm{SO}_n {\mathbb R} \to {\mathbb R}$, we have 
$$ Q(f) = 0 \, \Longleftrightarrow \, \exists (\rho, \rot) \in [0,\infty) \times \mathrm{SO}_n {\mathbb R} \mbox{ such that } f = \rho M_{\rot}. $$
\label{lem:equi}
\end{lemma}

\noindent
\textbf{Proof.} The left-to-right implication is clear. Conversely, let $f = \rho M_{\rot}$. We show that $\rho_f = \rho$ and $\rot_f = \rot$. The first equality is clear since $M_{\rot}$ is a probability density. Then, by Lemma \ref{lem:int_A_VM_dA}, we have $J_f = \rho \, c_1 \rot$ and since $\rho \, c_1 \geq 0$, by the uniqueness of the polar decomposition, $\rot$ is the orthogonal factor in the polar decomposition of $J_f$ and thus, we get $\rot_f = \rot$, which shows that $Q(f)=0$. \endproof

\begin{proposition}
The functions $\rho$ and $\rot$ involved in \eqref{eq:cvgce} satisfy the following equations: 
\begin{eqnarray}
&&\hspace{-1cm}
\int_{\mathrm{SO}_n {\mathbb R}} (\partial_t + A e_1 \cdot \nabla_x) (\rho M_{\rot}) \, dA = 0, \label{eq:mass} \\
&&\hspace{-1cm}
\int_{\mathrm{SO}_n {\mathbb R}} (\partial_t + A e_1 \cdot \nabla_x) (\rho M_{\rot}) \, (A \rot^T - \rot A^T) \, dA = 0, \label{eq:orient}
\end{eqnarray}
\label{prop:cvgce}
\end{proposition}

\noindent
\textbf{Proof.} We clearly have 
\begin{equation} 
\int_{\mathrm{SO}_n {\mathbb R}} Q(f) \, dA = 0, \, \mbox{ for all functions } \, f. 
\label{eq:1isaCI}
\end{equation}
Thus, integrating \eqref{eq:BGK} with respect to $A$ we get 
$$ \int_{\mathrm{SO}_n {\mathbb R}} (\partial_t + A e_1 \cdot \nabla_x) f^\varepsilon \, dA = 0. $$
Letting $\varepsilon \to 0$ and using \eqref{eq:cvgce} leads to \eqref{eq:mass}. 

Then, we remark that 
$$ \int_{\mathrm{SO}_n {\mathbb R}} Q(f) \, (A \rot_f^T - \rot_f A^T) \, dA = 0, \, \mbox{ for all functions } \, f. $$ 
Indeed, by Lemma \ref{lem:int_A_VM_dA}, we have 
$$ \int_{\mathrm{SO}_n {\mathbb R}} \rho_f M_{\rot_f} \, (A \rot_f^T - \rot_f A^T) \, dA = c_1 \rho_f (\rot_f \rot_f^T - \rot_f \rot_f^T) = 0. $$
Besides, by definition of the polar decomposition, there exists a symmetric matrix $S$ such that $J_f \rot_f^T =  S$. So, 
$$ \int_{\mathrm{SO}_n {\mathbb R}} f \, (A \rot_f^T - \rot_f A^T) \, dA = J_f \rot_f^T - \rot_f J_f^T= S - S^T = 0. $$
Thus, multiplying \eqref{eq:BGK} by $(A \rot_{f^\varepsilon}^T - \rot_{f^\varepsilon} A^T)$ and integrating with respect to $A$ we get 
$$ \int_{\mathrm{SO}_n {\mathbb R}} (\partial_t + A e_1 \cdot \nabla_x) f^\varepsilon \, (A \rot_{f^\varepsilon}^T - \rot_{f^\varepsilon} A^T) \, dA = 0. $$
Taking the limit $\varepsilon \to 0$ yields \eqref{eq:orient}. \endproof

\begin{remark}
In classical kinetic theory \cite{cercignani2013mathematical}, a key concept is that of ``collision invariant (CI)'' which, in the present context, is a function $\psi (A)$ such that
$$ \int_{\mathrm{SO}_n {\mathbb R}} Q(f) \, \psi \,  dA = 0, \quad \forall f. $$
It expresses that the quantity $\psi$ is conserved during collisions. The derivation of \eqref{eq:mass} examplifies how CI can be used to derive macroscopic equations. Indeed, Eq. \eqref{eq:1isaCI} expresses that constant functions are CI. However, these are the only CI and we do not have enough conservation relations to determine $\rho$ and $\rot$. In \cite{degond2008continuum}, a new ``generalized  collision invariant (GCI)'' concept was introduced to overcome this problem. The quantity $(A \rot^T - \rot A^T)$ used in the derivation of \eqref{eq:orient} is precisely a GCI. To make this concept more precise, for any $\rot \in \mathrm{SO}_n {\mathbb R}$, we introduce a linear collision operator
$$ {\mathcal Q} (f,\rot) = \rho_f M_\rot - f. $$
We note that $Q(f) = {\mathcal Q} (f,\rot_f)$. Then, for any fixed $\rot \in \mathrm{SO}_n {\mathbb R}$, we say that $\psi_\rot$ is a GCI associated with $\rot$ if and only if the following holds:
$$ \int_{\mathrm{SO}_n {\mathbb R}} {\mathcal Q} (f,\rot) \, \psi_\rot \,  dA = 0, \quad \forall f \textrm{ such that } J_f \rot^T \textrm{ is symmetric}, $$
and we immediately see that $(A \rot^T - \rot A^T)$ satisfies this requirement. Finally, we have 
\begin{equation} 
\int_{\mathrm{SO}_n {\mathbb R}} Q(f) \, \psi_{\rot_f} \, dA = 0, 
\label{eq:GCI}
\end{equation}
by the fact that $J_f \rot_f^T$ is symmetric. So, multiplying the kinetic equation by $\psi_{\rot_{f^\varepsilon}}$ and integrating with respect to $A$ cancels the $1/\varepsilon$ singularity and allows to pass to the limit $\varepsilon \to 0$ like in the end of the proof of Prop. \ref{prop:cvgce}. Eq. \eqref{eq:GCI} shows that $\psi_{\rot_f}$ satisfies a conservation relation like a classical CI, except that it depends on $f$ (a classical CI is independent of $f$). Because of the very simple relaxation form of the collision operator, the determination of the GCI is explicit. In more complex cases, the GCI are not explicit. For instance, in the Fokker-Planck case, they require the inversion of an elliptic operator~\cite{degond2017new, Degond_etal_MMS18}. 
\label{rem:gci}
\end{remark}

\subsection{The fluid model}
\label{sec_fluid}

We are left to re-write Eqs. \eqref{eq:mass}, \eqref{eq:orient} into a nice differential system for $\rho$ and $\rot$. This is easy for \eqref{eq:mass} as shown by the 

\begin{lemma}
Eq. \eqref{eq:mass} is equivalent to the continuity equation \eqref{eq:mass_2} where $c_1$ is given by \eqref{eq:int_A_VM_dA}. 
\label{lem:mass}
\end{lemma}

\noindent
\textbf{Proof.} Since $A e_1$ does not depend on $x$, we can move it inside the space derivative in \eqref{eq:mass} and then, we can interchange the time and space derivatives with the integral over $A$. Using that $M_\rot$ is a probability density and \eqref{eq:int_A_VM_dA}, we get \eqref{eq:mass_2}. \endproof

Before working out \eqref{eq:orient}, we recall the following definitions and facts: Let ${\mathcal S}_n$ and ${\mathcal A}_n$ the sub-spaces of ${\mathcal M}_n$ consisting of symmetric and antisymmetric matrices.  These two spaces form orthogonal complements with respect to the inner product \eqref{eq:innerprod}: 
$${\mathcal A}_n \stackrel{\bot}{\oplus} {\mathcal S}_n = {\mathcal M}_n. $$
Let now $\rot \in \mathrm{SO}_n {\mathbb R}$ and define the tangent manifold $T_\rot$ to $\mathrm{SO}_n {\mathbb R}$ at $\rot$. We have 
$$T_\rot = \{ P \rot \, \, | \, \, P \in {\mathcal A}_n \}. $$
The orthogonal projection (with respect to the inner product \eqref{eq:innerprod}) of $A \in {\mathcal M}_n$ onto $T_\rot$ is given by 
$$  P_{T_\rot} A = \frac{A \rot^T - \rot A^T}{2} \, \rot. $$
Likewise, the orthogonal complement $T_\rot^\bot$ is given by 
$$ T_\rot^\bot = \{ S \rot \, \, | \, \, S \in {\mathcal S}_n \}, $$
and the orthogonal projection of $A$ onto $T_\rot^\bot$ is  
$$ P_{T_\rot^\bot} A = \frac{A \rot^T + \rot A^T}{2} \, \rot. $$
For a given $A \in \mathrm{SO}_n {\mathbb R}$, we will need the derivative of the function $\mathrm{SO}_n {\mathbb R} \to {\mathbb R}$, $\rot \mapsto M_{\rot}(A)$ at $\rot$. This is a linear map $d_\rot M_{\rot} (A)$ from $T_\rot$ to ${\mathbb R}$ given for any $Q \in T_\rot$ by
\begin{equation} 
d_\rot M_{\rot} (A) (Q) = \kappa \, M_{\rot} (A) \, A \cdot Q = \kappa \, M_{\rot} (A) \, P_{T_\rot} A \cdot Q, 
\label{eq:deriv_VM}
\end{equation}
We will finally need the following lemma:

\begin{lemma} 
Let $F$: ${\mathcal M}_n^2 \to {\mathbb R}$ odd with respect to the first argument. Let $g$ : $\mathrm{SO}_n {\mathbb R} \to {\mathbb R}$ invariant by transposition ($g(A^T) = g(A)$, $\forall A \in \mathrm{SO}_n {\mathbb R}$). Then, for any $\rot \in \mathrm{SO}_n {\mathbb R}$, we have 
$$ \int_{\mathrm{SO}_n {\mathbb R}} F(P_{T_\rot} A, P_{T_\rot^\bot} A) \, g(A \rot^T) \, dA = 0.$$
\label{lem:oddness} 
\end{lemma}

\noindent
\textbf{Proof.} Changing variable to $A' = \rot A^T \rot$ in the integral changes the first argument of $F$ in its opposite while the second argument is unchanged. The factor involving $g$ is unchanged due to the invariance of $g$ by transposition and the Haar measure is unchanged due to its invariance by left and right multiplication and by transposition. Since $F$ is odd with respect to the first argument, the integral must be equal to zero. \endproof.

With these preliminaries, we remark that, after multiplication on the right by $\rot/2$,  \eqref{eq:orient} can be rewritten: 
\begin{equation}
 X := \int_{\mathrm{SO}_n {\mathbb R}} (\partial_t + A e_1 \cdot \nabla_x) (\rho M_{\rot}(A) ) \, P_{T_\rot} A \, dA = 0.
\label{eq:orient_2}
\end{equation}
Now, we have the 

\begin{lemma} 
We have $X = X_1 + X_2$, where 
\begin{eqnarray*}  
X_1 &=& \int_{\mathrm{SO}_n {\mathbb R}} \big[ (P_{T_\rot} A) e_1 \cdot \nabla_x \rho + \kappa \rho P_{T_\rot} A \cdot \partial_t \rot \big] P_{T_\rot} A \, M_{\rot}(A) \, dA, \\
X_2 &=& \kappa \rho \int_{\mathrm{SO}_n {\mathbb R}}  P_{T_\rot} A \cdot \big((P_{T_\rot^\bot} A) e_1 \cdot \nabla_x \big) \rot  \,  P_{T_\rot} A \, M_{\rot}(A) \, dA.
\end{eqnarray*}
\label{lem:decompX}
\end{lemma}

\noindent
\textbf{Proof.} Using \eqref{eq:deriv_VM}, we get
$$ (\partial_t + A e_1 \cdot\nabla_x) (\rho M_{\rot}) = M_{\rot} \big[ (\partial_t + A e_1 \cdot\nabla_x) \rho + \kappa \rho  P_{T_\rot} A \cdot [ (\partial_t + A e_1 \cdot\nabla_x) \rot] \big].$$
Now, using that $A = P_{T_\rot} A + P_{T_\rot^\bot} A$, we get $X = X_o + X_e$ with 
\begin{eqnarray*}  
X_o &=& \int_{\mathrm{SO}_n {\mathbb R}} \big[ \partial_t \rho + (P_{T_\rot^\bot} A) e_1 \cdot \nabla_x \rho + \kappa \rho \, P_{T_\rot} A \cdot \big( (P_{T_\rot} A) e_1 \cdot \nabla_x \big) \rot \big] P_{T_\rot} A \, M_{\rot}(A) \, dA, \\
X_e &=& \int_{\mathrm{SO}_n {\mathbb R}} \big[ (P_{T_\rot} A) e_1 \cdot \nabla_x \rho + \kappa \rho \, P_{T_\rot} A \cdot \partial_t \rot + \kappa \rho \, P_{T_\rot} A \cdot \big( (P_{T_\rot^\bot} A) e_1 \cdot \nabla_x \big) \rot \big] P_{T_\rot} A \, M_{\rot}(A) \, dA.
\end{eqnarray*}
Now, $X_o = 0$ thanks to Lemma \ref{lem:oddness} and $X_e = X_1 + X_2$. \endproof

\medskip
From $\rot$, we construct a matrix ${\mathbb P}$ and a rank-4 tensor ${\mathbb T}$ as follows: 
\begin{eqnarray}
{\mathbb P} &=& \frac{1}{2} \big( (\nabla_x \rho \otimes e_1) \, \rot^T - \rot \, (\nabla_x \rho \otimes e_1)^T \big) + \kappa \rho \, \partial_t \rot \, \rot^T, \label{eq:defP} \\
{\mathbb T}_{ijk \ell} &=& \frac{1}{2} \big( \rot_{k1} \, \rot_{jm} \, \partial_{x_\ell} \rot_{im}  + \rot_{\ell 1} \, \rot_{jm} \, \partial_{x_k} \rot_{im} \big), \label{eq:defT}
\end{eqnarray}
where $\otimes$ denotes the tensor product and tensor entries are defined with respect to the canonical basis of ${\mathbb R}^n$. Here and throughout the paper except in the appendices and unless otherwise specified, the repeated index summation convention is used. We note that ${\mathbb P}$ is antisymmetric while ${\mathbb T}$ is antisymmetric with respect to $(i,j)$ and symmetric with respect to $(k,\ell)$. In other words, 
\begin{eqnarray*}
{\mathbb P} &\in& {\mathcal A}_n, \\
{\mathbb T} &\in& {\mathcal A}_n \otimes {\mathcal S}_n. 
\end{eqnarray*}
The antisymmetry of ${\mathbb P}$ follows from $\partial_t \rot \in T_{\rot}$. The antisymmetry of ${\mathbb T}$ with respect to $(i,j)$ is a consequence of  
\begin{equation} 
\partial_{x_\ell} ( \rot_{jm} \rot_{im}) = \partial_{x_\ell} (\rot \rot^T)_{ij} = \partial_{x_\ell} \delta_{ij} = 0. 
\label{eq:antisym}
\end{equation}
The symmetry of ${\mathbb T}$ with respect to $(k,\ell)$ is by construction. We introduce the following maps:
\begin{eqnarray*}
L&:& {\mathcal A}_n \to {\mathcal A}_n, \quad P \mapsto L(P), \\
B&:& {\mathcal A}_n \times {\mathcal A}_n \to {\mathcal S}_n, \quad (P,Q) \mapsto B(P,Q), 
\end{eqnarray*}
given by 
\begin{eqnarray}
L(P) &=&  \int_{\mathrm{SO}_n {\mathbb R}} (A \cdot P)  \, \frac{A - A^T}{2} \, M_{\mathrm{Id}}(A) \, dA, 
\label{eq:defL}\\
B(P_1,P_2) &=& \int_{\mathrm{SO}_n {\mathbb R}} (A \cdot P_1) \, (A \cdot P_2) \, \frac{A + A^T}{2} \, M_{\mathrm{Id}}(A) \, dA. \label{eq:defB}
\end{eqnarray}
The maps $L$ and $B$ are respectively linear and bilinear symmetric. The following lemma gives alternate expressions for $X_1$ and $X_2$: 

\begin{lemma}
We have: 
\begin{equation}
X_1 = L({\mathbb P}) \, \rot, \label{eq:expressX1} 
\end{equation}
and for any matrix $P \in {\mathcal A}_n$, 
\begin{equation}
(X_2 \rot^T) \cdot P = \kappa \rho B_{k \ell} ({\mathbb T}_{\cdot \cdot k \ell}, P), \label{expressX2} 
\end{equation}
where ${\mathbb T}_{\cdot \cdot k \ell}$ stands for the antisymmetric matrix $({\mathbb T}_{\cdot \cdot k \ell})_{ij} = {\mathbb T}_{ijk \ell}$ and $B_{k \ell}$ is the $(k, \ell)$ entry of the symmetric matrix $B(\cdot,\cdot)$.  
\label{lem:expressX1X2}
\end{lemma}

\noindent
\textbf{Proof.}
For any $A \in {\mathcal M}_n$ we have $ Ae_1 \cdot \nabla_x \rho = A \cdot (\nabla_x \rho \otimes e_1)$. Thus, with \eqref{eq:defP}, we have: 
\begin{eqnarray*}  
X_1 &=& \int_{\mathrm{SO}_n {\mathbb R}} P_{T_\rot} A \cdot \big[ \nabla_x \rho \otimes e_1 + \kappa \rho \partial_t \rot \big] \, P_{T_\rot} A \, M_{\rot}(A) \, dA \\
&=& \int_{\mathrm{SO}_n {\mathbb R}} P_{T_\rot} A \cdot \big[ P_{T_\rot} (\nabla_x \rho \otimes e_1) + \kappa \rho \partial_t \rot \big] \, P_{T_\rot} A \, M_{\rot}(A) \, dA \\
&=& \int_{\mathrm{SO}_n {\mathbb R}} P_{T_\rot} A \cdot ({\mathbb P} \rot) \,  P_{T_\rot} A \, M_{\rot}(A) \, dA , 
\end{eqnarray*}
and by the invariance of the matrix  inner product by multiplication by rotations on the right and the change of variable $A' = A \rot^T$, we get expression \eqref{eq:expressX1} for $X_1$. 

We now turn to $X_2$. First, using the change of variable $A' = A \rot^T$ and the same invariance of the matrix inner product, we can write 
$$
X_2 =  \frac{\kappa \rho}{8} \int_{\mathrm{SO}_n {\mathbb R}}   (A - A^T) \cdot {\mathcal D}(\rot, (A+A^T) \rot e_1) \, (A - A^T) \,  M_{\mathrm{Id}}(A) \, dA \, \rot ,
$$
where, for a vector $w \in {\mathbb R}^n$, we let
$$ {\mathcal D}(\rot, w) = (w \cdot \nabla_x) \rot \, \rot^T \in {\mathcal A}_n. $$
We have the following lemma, the proof of which is deferred to the end of the present proof. 

\begin{lemma}
We have 
$$ {\mathcal D}(\rot, w)_{ij} = w_\ell \partial_{x_\ell} \rot_{i m} \, \rot_{j m}. $$
\label{lem:expresscalD}
\end{lemma}

\noindent
Thanks to this lemma and to \eqref{eq:defT}, we have:
\begin{eqnarray*}
{\mathcal D}(\rot, (A+A^T) \rot e_1)_{ij} &=&  \big((A+A^T) \rot e_1 \big)_\ell \,  \partial_{x_\ell} \rot_{i m} \, \rot_{j m} \\
&=& (A_{\ell k} + A_{k \ell}) \, \rot_{k1} \, \partial_{x_\ell} \rot_{im} \, \rot_{jm}  = (A_{k \ell} + A_{\ell k}) {\mathbb T}_{ijk\ell}, 
\end{eqnarray*}
and thus
$$
X_2 \, \rot^T = \frac{\kappa \rho}{8} {\mathbb T}_{ijk\ell} \, \Big( \int_{\mathrm{SO}_n {\mathbb R}} (A_{ij} - A_{ji}) \,  (A_{k \ell} + A_{\ell k}) \, (A - A^T) \, M_{\mathrm{Id}}(A) \, dA \Big) . 
$$
Taking the matrix inner product with $P$ for any $P \in {\mathcal A}_n$, we get \eqref{expressX2}. \endproof

\noindent 
\textbf{Proof of Lemma \ref{lem:expresscalD}.}
Define the matrix $\dot \rot_w$ by $(\dot \rot_w)_{ij} = (w \cdot \nabla_x) \rot_{ij}$. Then, we have $(w \cdot \nabla_x) \rot = P_{T_\rot} \dot \rot_w$ and so, with \eqref{eq:antisym}, 
\begin{eqnarray*} 
{\mathcal D}(\rot, w)_{ij} &=& \frac{1}{2} (\dot \rot_w \, \rot^T - \rot \, \dot \rot_w^T)_{ij} \\
&=& \frac{1}{2} \big( (w \cdot \nabla_x) \rot_{im} \, \rot_{jm} - \rot_{im} \,  (w \cdot \nabla_x) \rot_{jm} \big) = (w \cdot \nabla_x) \rot_{im} \, \rot_{jm},   
\end{eqnarray*}
which shows the lemma. \endproof

Now, we express the mappings $L$ and $B$ in the following Lemma, whose proof is given in Sections \ref{sec:proof:expressLBi} and \ref{sec:proof:expressLBii}.

\begin{lemma}
(i) We have
\begin{equation} 
L(P) = C_2 \, P, \, \,  \forall P \in {\mathcal A}_n, \, \, \textrm{ with } \, \, C_2 = \frac{1}{n-1} \Big(1- \big \langle \frac{\mathrm{Tr} \, A^2}{n}  \big \rangle_{\exp (\kappa \mathrm{Tr} A )} \Big). 
\label{eq:expressL}
\end{equation}

\noindent
(ii) We have
\begin{equation} 
B(P,Q) =  C_3 \, \mathrm{Tr} \, (PQ) \, \mathrm{Id} + C_4 \, \Big( \frac{PQ + QP}{2} - \frac{1}{n} \mathrm{Tr} \, (PQ) \, \mathrm{Id} \Big),  \, \,  \forall P, \, Q \in {\mathcal A}_n,
\label{eq:expressB}
\end{equation}
with 
\begin{eqnarray}
C_3 &=& \frac{1}{n-1} \Big\langle \Big(\frac{\mathrm{Tr} A^2}{n}  - 1 \Big) \, \frac{\mathrm{Tr} A}{n}   \Big \rangle_{\exp (\kappa \mathrm{Tr} A )}, \label{eq:expressc3} \\
C_4 &=& \frac{2n}{(n-1)(n-2)(n+2)} \Big\langle  \frac{\mathrm{Tr} A^3}{n} - 2 \frac{\mathrm{Tr} A}{n}  \, \frac{\mathrm{Tr} A^2}{n} + \frac{\mathrm{Tr} A}{n} \Big\rangle_{\exp (\kappa \mathrm{Tr} A )}. \label{eq:expressc4}
\end{eqnarray}
\label{lem:expressLB}
\end{lemma}

\begin{remark}
The proof presented in Sections \ref{sec:proof:expressLBi} and \ref{sec:proof:expressLBii} is based on representation theory. But this lemma can be proved using elementary algebra: Eq. \eqref{eq:expressL} is proved in \cite[Lemma 3.4]{Degond_eal_JNLS20} in dimension $n\geq3$, $n\ne4$; Eq.  \eqref{eq:expressB} can be proved with a similar approach as outlined in Section 9, Remark \ref{rem:elementary_proof}. 
\label{rem:elementary}
\end{remark}

From this, we can prove the 

\begin{lemma}
Eq. \eqref{eq:orient} is equivalent to \eqref{eq:orient_3} with $c_3$ given by \eqref{eq:express_c3} and $c_2$ and $c_4$ by:
\begin{eqnarray} 
\hspace{-0.5cm} c_2 &=& \frac{1}{(n-2)(n+2)} \, \frac{\Big\langle 2 \frac{\mathrm{Tr} \, A^3}{n} - n^2 \, \frac{\mathrm{Tr} \, A}{n} \, \frac{\mathrm{Tr} \, A^2}{n} + (n^2 - 2) \, \frac{\mathrm{Tr} \, A}{n} \Big\rangle_{e^{\kappa \mathrm{Tr} \, A}}}{\Big \langle 1- \frac{\mathrm{Tr} \, A^2}{n} \Big\rangle_{e^{\kappa \mathrm{Tr} \, A}}}, \label{eq:express_c2} \\
\hspace{-0.5cm} c_4 &=& \frac{n}{2(n-2)(n+2)}  \, \frac{\Big\langle \frac{\mathrm{Tr} \, A^3}{n} - 2 \, \frac{\mathrm{Tr} \, A}{n} \, \frac{\mathrm{Tr} \, A^2}{n} + \frac{\mathrm{Tr} \, A}{n} \Big\rangle_{e^{\kappa \mathrm{Tr} \, A}}}{\Big \langle 1- \frac{\mathrm{Tr} \, A^2}{n} \Big\rangle_{e^{\kappa \mathrm{Tr} \, A}}}. \label{eq:express_c4}
\end{eqnarray}
\label{lem:orient}
\end{lemma}

\noindent
\textbf{Proof.} 
We first note that $(\Omega_k)_\ell = \rot_{\ell k}$. Thus
$$ (\nabla_x \rho \otimes e_1) \, \rot^T - \rot \, (\nabla_x \rho \otimes e_1)^T  = \nabla_x \rho \wedge \Omega_1. $$
Thus, introducing \eqref{eq:expressL} into \eqref{eq:expressX1} and using \eqref{eq:defP}, we get 
\begin{equation}
 X_1 = C_2 \, {\mathbb P} \, \rot = \kappa C_2 \, \big[ \rho \partial_t \rot + c_3 \, (\nabla_x \rho \wedge \Omega_1) \, \rot \big]. 
\label{eq:expressX1_2}
\end{equation}

Now, combining \eqref{eq:expressB} and \eqref{expressX2}, we get for any $P \in {\mathcal A}_n$, 
\begin{equation}
(X_2 \rot^T) \cdot P = \kappa \rho \, \Big[ \Big( C_3 - \frac{C_4}{n} \Big) \, \mathrm{Tr} ({\mathbb T}_{\cdot \cdot k k} P)  + \frac{C_4}{2} \, \big( {\mathbb T}_{\cdot \cdot k \ell} P + P {\mathbb T}_{\cdot \cdot k \ell} \big)_{k \ell} \Big] . \label{eq:X2.PT}
\end{equation}
Using that $P$ is antisymmetric, we have $\mathrm{Tr} ({\mathbb T}_{\cdot \cdot k k} P) =  - {\mathbb T}_{\cdot \cdot k k} \cdot  P$ and, using \eqref{eq:defT} together with Lemma \ref{lem:expresscalD}, we get
$$({\mathbb T}_{\cdot \cdot k k})_{ij}  = \rot_{k1} \, \partial_{x_k} \rot_{im} \, \rot_{jm} = \big( (\Omega_1 \cdot \nabla_x) \rot \, \rot^T \big)_{ij}.  $$
This yields
\begin{equation}  
\mathrm{Tr} ({\mathbb T}_{\cdot \cdot k k} P) = - \big( (\Omega_1 \cdot \nabla_x) \rot \, \rot^T \big) \cdot P. 
\label{eq:TrT..}
\end{equation}

Similarly, we compute
$$ \big( {\mathbb T}_{\cdot \cdot k \ell} P + P {\mathbb T}_{\cdot \cdot k \ell} \big)_{k \ell} = {\mathbb T}_{k m k \ell} P_{m \ell} + P_{km} {\mathbb T}_{m \ell k \ell}. $$
But we have 
$$ {\mathbb T}_{m \ell k \ell} P_{km} = - {\mathbb T}_{\ell m \ell k} P_{km} = -{\mathbb T}_{k \ell k m} P_{m \ell}, $$
where the first equality exploits the fact that ${\mathbb T}_{ijk\ell}$ is antisymmetric with respect to $(i,j)$ and symmetric with respect to $(k,\ell)$, and the second one is just the circular permutation $k \to m \to \ell \to k$. Thus:
\begin{equation} 
\big( {\mathbb T}_{\cdot \cdot k \ell} P + P {\mathbb T}_{\cdot \cdot k \ell} \big)_{k \ell} =({\mathbb T}_{k m k \ell} - {\mathbb T}_{k \ell k m}) P_{m \ell} = {\mathbb V}_{m \ell} \, P_{m \ell} = {\mathbb V} \cdot P, 
\label{eq:calculT0}
\end{equation}
with 
$$ {\mathbb V}_{m \ell} = {\mathbb T}_{k m k \ell} - {\mathbb T}_{k \ell k m}. $$
Using \eqref{eq:defT}, we have 
\begin{equation}
2 {\mathbb V}_{m \ell} = \rot_{k1} \, \rot_{mp} \, \partial_{x_\ell} \rot_{kp} + \rot_{\ell 1} \, \rot_{mp} \, \partial_{x_k} \rot_{kp} 
- \rot_{k1} \, \rot_{\ell p} \, \partial_{x_m} \rot_{kp} - \rot_{m 1} \, \rot_{\ell p} \, \partial_{x_k} \rot_{kp}. \label{eq:calculT}
\end{equation}
The second and fourth terms of \eqref{eq:calculT} give
\begin{eqnarray*} 
\rot_{\ell 1} \, \rot_{mp} \, \partial_{x_k} \rot_{kp} - \rot_{m 1} \, \rot_{\ell p} \, \partial_{x_k} \rot_{kp} &=& \big[ (\Omega_1)_\ell \, (\Omega_p)_m - (\Omega_1)_m \, (\Omega_p)_\ell \big] \, \nabla_x \cdot \Omega_p \\
&=& (\Omega_p \wedge \Omega_1)_{m \ell} \, \nabla_x \cdot \Omega_p = (r \wedge \Omega_1 )_{m \ell}. 
\end{eqnarray*}
Using a similar computation as \eqref{eq:antisym}, the first and third terms of \eqref{eq:calculT} lead to
\begin{eqnarray*}
\rot_{k1} \, \rot_{mp} \, \partial_{x_\ell} \rot_{kp} - \rot_{k1} \, \rot_{\ell p} \, \partial_{x_m} \rot_{kp} &=& - \rot_{k1} \, \rot_{kp} \, ( \partial_{x_\ell} \rot_{mp} - \partial_{x_m} \rot_{\ell p})  \\
&=& - (\partial_{x_\ell} (\Omega_1)_m - \partial_{x_m} (\Omega_1)_\ell) = (\nabla_x \wedge \Omega_1)_{m \ell}.
\end{eqnarray*}
Collecting these two terms, this gives 
\begin{equation}
{\mathbb V} = \frac{1}{2} \big( r \wedge \Omega_1 + \nabla_x \wedge \Omega_1 \big). 
\label{eq:expressV}
\end{equation}

Finally, collecting \eqref{eq:X2.PT}, \eqref{eq:TrT..}, \eqref{eq:calculT0} and \eqref{eq:expressV}, we get
$$ (X_2 \rot^T) \cdot P  = \kappa \rho \, \Big[ - \Big(C_3 - \frac{C_4}{n} \Big) \,  (\Omega_1 \cdot \nabla_x) \rot \, \rot^T   + \frac{C_4}{4} \, ( r \wedge \Omega_1 + \nabla_x \wedge \Omega_1 ) \Big] \cdot P. $$
Since the matrix inside the bracket is antisymmetric and this identity is valid for any $P \in {\mathcal A}_n$, we get
\begin{equation}
X_2  = \kappa C_2 \, \rho \, \Big[ c_2 \,  (\Omega_1 \cdot \nabla_x) \rot \,  + c_4 \, ( r \wedge \Omega_1 + \nabla_x \wedge \Omega_1 ) \, \rot \Big], 
\label{eq:expressX2_2}
\end{equation}
where 
\begin{equation} 
c_2 = - \frac{1}{C_2} \Big( C_3 - \frac{C_4}{n} \big), \qquad c_4 = \frac{C_4}{4 C_2}. 
\label{eq:constants}
\end{equation}
Now, collecting \eqref{eq:expressX1_2} and \eqref{eq:expressX2_2}, we get \eqref{eq:orient_3}, \eqref{eq:expressW}, \eqref{eq:expressF} with $c_3$ given by \eqref{eq:express_c3}. Finally, inserting \eqref{eq:expressL}, \eqref{eq:expressc3} and \eqref{eq:expressc4} into \eqref{eq:constants}, we get \eqref{eq:express_c2}, \eqref{eq:express_c4}, which ends the proof. \endproof

Finally, we show that $c_1$, $c_2$ and $c_4$ are given by the expressions \eqref{eq:express_c1_WIF}, \eqref{eq:express_c2_WIF} and \eqref{eq:express_c4_WIF} thanks to Weyl's integration formula (see Section \ref{sec:rep_theory}). 

\begin{lemma}\label{lemma:coefficientsWeyl}
Let $n \in {\mathbb N}$, $n\geq 3$ and let $p \in {\mathbb N}$ such that $n=2p$ or $n=2p+1$. We have $c_1$, $c_2$ and $c_4$ given by \eqref{eq:express_c1_WIF}, \eqref{eq:express_c2_WIF}, \eqref{eq:express_c4_WIF} respectively. 
\label{lem:constc2c4}
\end{lemma}

\noindent
\textbf{Proof.} We first remark that for any integer $k$, the function $A \mapsto \mathrm{Tr} A^k$ is a class function. So, we can apply Weyl's integration formula (see Section \ref{sec:rep_theory}) to all the integrals involved in formulas \eqref{eq:int_A_VM_dA}, \eqref{eq:express_c2}, \eqref{eq:express_c4}.  We note that 
$$ (R_{\theta_1,  \ldots , \theta_p})^k = R_{k \theta_1,  \ldots , k \theta_p}, \quad \forall k \in {\mathbb Z}, $$ 
where $R_{\theta_1 \ldots \theta_p}$ is defined by \eqref{eq:R2p} in the case of $\mathrm{SO}_{2p}{\mathbb R}$ and by \eqref{eq:R2p+1} in the case of $\mathrm{SO}_{2p+1}{\mathbb R}$. We have
$$ \mathrm{Tr} \, R_{\theta_1,  \ldots , \theta_p} = \left\{ \begin{array}{lll}
2(\cos  \theta_1 + \ldots + \cos  \theta_p) & \mbox{ if } & n=2p, \\
2(\cos  \theta_1 + \ldots + \cos  \theta_p) + 1& \mbox{ if } & n=2p+1,
\end{array} \right. 
$$
so that 
$$ \mathrm{Tr} \, (R_{\theta_1,  \ldots , \theta_p})^k  = \left\{ \begin{array}{lll}
C_{2p}^{(k)} & \mbox{ if } & n=2p, \\
C_{2p+1}^{(k)} & \mbox{ if } & n=2p+1. 
\end{array} \right. 
$$
Lemma \ref{lem:constc2c4} follows immediately. \endproof

Collecting Lemmas \ref{lem:mass}, \ref{lem:orient} and \ref{lem:constc2c4} shows Theorem \ref{thm:main}, which ends this section.

\setcounter{equation}{0}
\section{Conclusion and perspectives}
\label{sec_conclu}

In this paper, we have derived the hydrodynamic limit of a kinetic model of self-propelled agents interacting through body attitude coordination in arbitrary dimension $n \geq 3$. Previous literature was restricted to dimension $n=3$. In arbitrary dimension, the derivation uses Lie group representations and the Weyl integration formula. The obtained hydrodynamic model is structurally identical to that obtained in dimension $3$ (and referred to as the ``Self-Organized Hydrodynamics for Body orientation (SOHB)'') but the constants involved have expressions that depend on the dimension. Future work will be concerned with existence of solutions for the SOHB model, rigorous convergence from the kinetic to the SOHB model and derivation of explicit solutions. We will also investigate the situation where the particle dynamics are described by stochastic differential equations instead of PDMP as considered here. In this case, the resulting kinetic model involves a Fokker-Planck operator for which the generalized collision invariants are still unknown. As we have seen, knowing the expression of the GCI is crucial to get an explicit expression of the coefficients of the hydrodynamic model.  The numerical resolution of the SOHB model has not been undertaken yet and will require the design of appropriate numerical schemes. Finally, the qualitative properties of the solutions of the SOHB model, and particularly, their topology, need to be further investigated.

\newpage

\begin{center}
\bf\LARGE Appendices
\end{center}

In the forthcoming sections, we assume $n \geq 3$. The proofs contained in these appendices rely on results from representation theory. We start with recalling a few useful results from this theory. We refer to \cite{Fulton_Harris} for the terminology and notations. In all sections of this appendix, the repeated index summation convention is \textbf{not} used.

\setcounter{equation}{0}
\section{Short summary of useful results from representation theory}
\label{sec:rep_theory}

Let $G$ be a Lie group. A representation of $G$ on the  vector space $V = {\mathbb R}^n$ or ${\mathbb C}^n$ is a group morphism: $G \to \textrm{GL} (V)$ into the group of automorphisms of $V$. Likewise, if $\mathfrak{g}$ is a Lie algebra, a representation of $\mathfrak{g}$ is a map of Lie algebras $\mathfrak{g} \to \mathfrak{gl} (V)$, where $\mathfrak{gl} (V)$ is the space of endomorphisms of $V$. Here, we will be mostly concerned with the representations of $\mathrm{SO}_n {\mathbb R}$ which acts on ${\mathcal M}_n$ by conjugation, i.e.  the action of $R \in \mathrm{SO}_n {\mathbb R}$ sends $M \in {\mathcal M}_n$ to $R M R^T$. The reason is that our objects of study have remarkable transforms under this action. There is a strong connection between representations of a Lie group and representations of its Lie algebra. Lie algebras have a rich structure and one starts by constructing representations of a Lie algebra before lifting them to the Lie groups that have this Lie algebra in common. We note that the base field is a representation of $G$ or $\mathfrak{g}$. For instance, it sends all the elements of the group to the identity. This is called the trivial representation. If $G$ is a matrix group, i.e. a subgroup of $\textrm{GL} (V)$, then the mapping $\rho$: $G \to \textrm{GL} (V)$ such that $\rho(g) = g$ is also a representation called the standard representation. 

A representation is said irreducible if it has no proper subspace which is left invariant by the representation. In good cases (which include those we will consider), any representation can be decomposed into the direct sum of irreducible representations, making irreducible representations the building blocks of the theory. The reason  why irreducible representations are so appealing is the so called Schur Lemma: 
\begin{lemma}[Schur Lemma] Let $V$ and $W$ be two irreducible complex representations of a group $G$ and let $T$: $V \mapsto W$ be a map of representations, i.e. a linear map which commutes with the representations (one also says, a map which intertwins the representations). Then, there exists $C \in {\mathbb C}$ such that $T = C \, \mathrm{Id}$. Furthermore, $C=0$ if the two representations are not isomorphic. 
\label{lem:Schur}
\end{lemma}
One should be careful that the result does not hold in these terms for real representations: if the two representations are not isomorphic $T$ is still $0$ but if the two representations are isomorphic, then $T$ is an isomorphism, but we cannot say that $T = C \, \mathrm{Id}$, except in some special cases. Anyhow,  intertwinning maps of irreducible representations have a very simple structure and we want to exploit this structure in the results below. So, for a given representation, we want to find its decomposition in irreducible representations. 

The theory starts to construct the  finite-dimensional irreducible representations of the Lie algebra $\mathfrak{sl}_n{\mathbb C}$ as subspaces of tensor products $V^{\otimes d}$ of the standard representation $V = {\mathbb C}^n$. These representations are in bijective correspondence to conjugacy classes of the symmetric group $\mathfrak{S}_d$, themselves in bijective correspondence to partitions $\lambda = (\lambda_1, \ldots, \lambda_{n-1})$ of $d$,  i.e. such that $\lambda_1 \geq \lambda_2 \geq \ldots \geq \lambda_{n-1} \geq 0$ and $\lambda_1 + \ldots + \lambda_{n-1} = d$. The irreducible representation associated with $\lambda$ is called the Schur functor (or Weyl module) ${\mathbb S}_{\lambda}(V)$ \cite[\S 15.3]{Fulton_Harris}. 

Among remarkable irreducible representations of  $\mathfrak{sl}_n{\mathbb C}$ are the symmetric and exterior powers of $V$ \cite[Appendix B]{Fulton_Harris}. The symmetric power $\textrm{Sym}^d(V)$ is the space of  symmetric tensors. For $v_1, \ldots, v_d \in V$, we denote by $v_1 \circ \ldots \circ v_d = \sum_{\sigma \in \mathfrak{S}_d} v_{\sigma(1)} \otimes \ldots \otimes v_{\sigma(d)}$, the symmetric product of $v_1, \ldots, v_d$. We have $\textrm{Sym}^d(V) = \textrm{Span} \{ v_1 \circ \ldots \circ v_d \,| \, v_1, \ldots, v_d \in V \} \subset V^{\otimes d}$. Likewise, the exterior power $\Lambda^d(V)$ is the space of antisymmetric tensors. For $v_1, \ldots, v_d \in V$, we denote by $v_1 \wedge \ldots \wedge v_d = \sum_{\sigma \in \mathfrak{S}_d} \varepsilon(\sigma) v_{\sigma(1)} \otimes \ldots \otimes v_{\sigma(d)}$, the exterior product of $v_1, \ldots, v_d$, where $\varepsilon(\sigma)$ is the signature of the permutation $\sigma$. We have $\Lambda^d(V) = \textrm{Span} \{ v_1 \wedge \ldots \wedge v_d \,| \, v_1, \ldots, v_d \in V \} \subset V^{\otimes d}$. $\textrm{Sym}^d(V)$ and $\Lambda^d(V)$ are irreducible representations of $\mathfrak{sl}_n{\mathbb C}$ which correspond to ${\mathbb S}_{\lambda}(V)$ for the partitions $\lambda = (d,0, \ldots, 0)$ (abbreviated by $\lambda = d$)  and $\lambda = (1, \ldots , 1)$ ($d$ times), \cite[\S 15.3]{Fulton_Harris}. 

It is possible to pass from $\mathfrak{sl}_n{\mathbb C}$ to $\mathfrak{so}_{n}{\mathbb C}$ (for $n \geq 3)$ by means of the Weyl construction \cite[\S 19.5]{Fulton_Harris}. Let $I=(p,q)$ with $p<q$ be any pair of integers in $\{1, \ldots, d \}$. The contraction $\Phi_I$ is the mapping 
$$ \Phi_I: \left\{ \begin{array}{lll} 
V^{\otimes d} & \rightarrow & V^{\otimes (d-2)}  \\
v_1 \otimes \ldots \otimes v_d & \to & (v_p \cdot v_q) \, v_1 \otimes \ldots \otimes \hat v_p \otimes \ldots \otimes \hat v_q \otimes \ldots \otimes v_d , 
\end{array} \right.
$$
where the hat means that the corresponding factor is removed. $V^{[d]}$ denotes the intersection of the kernels of such contractions over all pairs $I$ of indices. Then, ${\mathbb S}_{[\lambda]}(V) = {\mathbb S}_{\lambda}(V) \cap V^{[d]}$, when it is not $\{0\}$,  is an irreducible representation of $\mathrm{O}_n{\mathbb C}$. For two associated partitions in the sense of Weyl, $\lambda$ and $\mu$, ${\mathbb S}_{[\lambda]}(V)$ and ${\mathbb S}_{[\mu]}(V)$ are isomorphic as irreducible representations of $\mathfrak{so}_{n}{\mathbb C}$ (see \cite[\S 19.5]{Fulton_Harris} for the definition of associated partitions). Furthermore, if $\lambda$ is associated to itself, ${\mathbb S}_{[\lambda]}(V)$ is not irreducible on $\mathfrak{so}_{n}{\mathbb C}$ but decomposes into the sum of two non-isomorphic irreducible representations of the same dimension. We will encounter this case in Section \ref{sec:proof:expressLBi} with $\Lambda^2({\mathbb C}^4)$ and in Section \ref{sec:proof:expressLBii} with $\Lambda^4({\mathbb C}^8)$. 

The representations ${\mathbb S}_{[\lambda]}(V)$ when they are irreducible on $\mathfrak{so}_{n}{\mathbb C}$, lift to irreducible representations of $\mathrm{SO}_n{\mathbb C}$ \cite[Proposition 23.13 (iii)]{Fulton_Harris}. Note that not all irreducible representations of $\mathfrak{so}_{n}{\mathbb C}$ are obtained this way: to complete the list one needs to introduce the spin representations but they do not lift to  representations of $\mathrm{SO}_n{\mathbb C}$ \cite[Proposition 23.13 (iii)]{Fulton_Harris} and will be ignored here. The complex irreducible representations of $\mathrm{SO}_n{\mathbb C}$ and $\mathrm{SO}_n{\mathbb R}$ are the same (see \cite[\S 26.1, Section ``Real groups'']{Fulton_Harris} for details). Finally complex irreducible representations of $\mathrm{SO}_n{\mathbb R}$ will give rise to real ones (in other words, these complex representations are complexifications of real representations) in mostly all cases. The only troublesome cases are $n=2p$ even with either $p$ odd and $\lambda_n \not = 0$ or $p \equiv 2 \, \textrm{mod} \, 4$ and $\lambda_{n-1}$ odd \cite[Prop. 26.26 and 26.27]{Fulton_Harris} and we will not meet them. 

Let now $G$ be a compact Lie group (such as $\mathrm{SO}_n{\mathbb R}$) endowed with its Haar measure $dg$ and let $\rho$: $V \to \mathrm{GL}(V)$ be a representation of $G$. The character of $V$, denoted  by $\chi_V$, is the mapping $G \ni g \mapsto \chi_V(g) = \mathrm{Tr} \rho(g) \in {\mathbb C}$. If $V$ and $W$ are irreducible, we have Schur's orthogonality relation \cite[\S 26.2]{Fulton_Harris}: 
\begin{equation} 
\int_G \chi_V(g) \, \bar \chi_W(g) \, dg = \left\{ \begin{array}{ll} 1 & \textrm{ if }  V \textrm{ and } W \textrm{ are isomorphic}, \\
0 & \textrm{ otherwise}. \end{array} \right. 
\label{eq:Schur_ortho}
\end{equation}

The final result of representation theory that we will need is the Weyl integration formula. For $\mathrm{SO}_n{\mathbb R}$, its statement is given in \cite[Theorems IX.9.4 \& IX.9.5]{Simon}. We first introduce some notations. For $\theta \in {\mathbb R}$, we define the planar rotation matrix $R_\theta$ by
$$ R_\theta = \left( \begin{array}{rr} \cos \theta & - \sin \theta \\ \sin  \theta &  \cos \theta \end{array} \right). $$
For $(\theta_1, \ldots , \theta_p) \in {\mathbb R}^p$, $R_{\theta_1, \ldots ,\theta_p}$ denotes the following matrix defined by blocks:
\begin{itemize}
\item in the case $n = 2p$, $p \geq 2$, 
\begin{equation} R_{\theta_1, \ldots , \theta_p} = \left( \begin{array}{cccc} 
\scalebox{1.2}{$R_{\theta_1}$} &  &  & \scalebox{2.}{$0$} \\
& \scalebox{1.2}{$R_{\theta_2}$} &  & \\
&  &  \ddots  &  \\
\scalebox{2.}{$0$} &  &  &  \scalebox{1.2}{$R_{\theta_p}$}
\end{array} \right) \in \mathrm{SO}_{2p}{\mathbb R}, 
\label{eq:R2p}
\end{equation}
\item in the case $n = 2p+1$, $p \geq 1$, 
\begin{equation}
R_{\theta_1 , \ldots , \theta_p} = \left( \begin{array}{ccccc} 
\scalebox{1.2}{$R_{\theta_1}$} &  &  & \scalebox{2.}{$0$} & 0 \\
& \scalebox{1.2}{$R_{\theta_2}$} &  & & \vdots \\
&  &  \ddots  &  & \vdots \\
\scalebox{2.}{$0$} &  &  &  \scalebox{1.2}{$R_{\theta_p}$} & 0 \\
0  &  \ldots & \ldots & 0 & 1 
\end{array} \right) \in \mathrm{SO}_{2p+1}{\mathbb R}. 
\label{eq:R2p+1}
\end{equation}
\end{itemize}
We define a maximal torus $T$ of $\mathrm{SO}_{2p}{\mathbb R}$ or $\mathrm{SO}_{2p+1}{\mathbb R}$ by 
$$ T = \{ R_{\theta_1 \ldots \theta_p}  \, \, | \, \, (\theta_1, \ldots , \theta_p) \in {\mathbb R}^p \} .$$
The maximal torus $T$ is an abelian subgroup of $\mathrm{SO}_{2p}{\mathbb R}$ or $\mathrm{SO}_{2p+1}{\mathbb R}$ isomorphic to the $p$-dimensional torus. We recall that any element of $\mathrm{SO}_{2p}{\mathbb R}$ or $\mathrm{SO}_{2p+1}{\mathbb R}$ is conjugate to a (non-unique) element of $T$, i.e. $\forall A \in \mathrm{SO}_{2p}{\mathbb R}$, $\exists B \in T$, $\exists U \in \mathrm{SO}_{2p}{\mathbb R}$ such that $A = UBU^T$ (and similarly with $2p+1$). We also recall that a class function $f$ on a group $G$ is a function that only depends on the conjugation class, i.e. a function that takes the same value on two conjugate elements of the group.

\begin{proposition}[Weyl integration formula (\cite{Simon}, Theorems IX.9.4 \& IX.9.5)] $\mbox{}$

\noindent
 Let $n \in {\mathbb N}$, $n\geq 3$. Let $p \in {\mathbb N}$ such that $n=2p$ or $n=2p+1$. For any integrable class function $f$ on $\mathrm{SO}_n{\mathbb R}$, we have 
\begin{equation}
\int_{\mathrm{SO}_n{\mathbb R}} f(A) \, dA = \frac{1}{(2 \pi)^p} \int_{[0,2 \pi]^p} f(R_{\theta_1 \ldots \theta_p}) \, u_n (\theta_1, \ldots, \theta_p) \, d \theta_1 \, \ldots \, d \theta_p,
\label{eq:WIF}
\end{equation}
where $u_n$: ${\mathbb R}^p \to {\mathbb R}$ is defined by \eqref{eq:u2p} (in the case $n=2p$) or \eqref{eq:u2p+1} (if $n=2p+1$). 
\label{prop:WIF}
\end{proposition}

\begin{remark}
(i) There is a typo in \cite[Theorems IX.9.4]{Simon}. The normalization is not correct as can be realized by taking $f=1$ and $p=1$. It would correspond to taking the constant in \eqref{eq:u2p+1} equal to $\frac{2^{p(p-2)}}{p!}$. We have carefully redone the computation and \eqref{eq:u2p+1} can be easily checked for $p=1$ and $p=2$. \\
(ii) Taking $f = 1$ in \eqref{eq:WIF} leads to $(2 \pi)^{-p}\int_{[0,1]^p} u_n \, d \theta_1 \, \ldots \, d \theta_p =1$. A direct proof does not seem obvious. 
\end{remark}


\setcounter{equation}{0}
\section{Proof of Lemma \ref{lem:int_A_VM_dA}}
\label{sec:proof:int_A_VM_dA}

Changing variable to $A' = \rot^T A$ in the integral appearing in \eqref{eq:int_A_VM_dA} yields 
$$ \int_{\mathrm{SO}_n {\mathbb R}}  A \, M_{\rot}(A) \, dA = \rot \int_{\mathrm{SO}_n {\mathbb R}}  A \, M_{\mathrm{Id}}(A) \, dA. $$
We will prove that for any $g$: $\mathrm{SO}_n {\mathbb R} \to {\mathbb R}$, $A \mapsto g(A)$ invariant by conjugation $g(U A U^T) = g(A)$, $\forall A, \, U \in \mathrm{SO}_n {\mathbb R}$ (i.e. $g$ is a class function) and by transposition $g(A^T) = g(A)$, we have 
\begin{equation} 
\int_{\mathrm{SO}_n {\mathbb R}}  A \, g(A) \, dA = C_1 \, \mathrm{Id},  \, \, \textrm{ with } \, \, C_1 = \frac{1}{n} \int_{\mathrm{SO}_n {\mathbb R}}  \mathrm{Tr} A \, g(A) \, dA . 
\label{eq:intAgdA}
\end{equation}
Obviously, $g(A) =  M_{\mathrm{Id}}(A) = Z^{-1} \exp (\mathrm{Tr} A)$ is a class function invariant by transposition and \eqref{eq:intAgdA} directly implies Lemma \ref{lem:int_A_VM_dA}. First, we remark that the second formula of \eqref{eq:intAgdA} is a direct consequence of the first one by taking the trace. We now show \eqref{eq:intAgdA}.

Using that $g$ and the Haar measure are invariant by transposition, \eqref{eq:intAgdA} is equivalent to saying that 
\begin{equation} 
\int_{\mathrm{SO}_n {\mathbb R}}  \big( \frac{A + A^T}{2} - \frac{\mathrm{Tr} A}{n} \, \mathrm{Id} \big) \, g(A) \, dA = 0.
\label{eq:intA+ATgdA}
\end{equation}
Let ${\mathcal S}_n^0$ be the subspace of ${\mathcal S}_n$ consisting of trace-free matrices and define the mapping $K$: ${\mathcal S}_n^0 \to {\mathbb R}$, $S \mapsto K(S)$ by 
$$ K(S) = \int_{\mathrm{SO}_n {\mathbb R}}  A \cdot S \, g(A) \, dA. $$
Eq. \eqref{eq:intA+ATgdA} is equivalent to saying that  $K=0$. Let $U \in \mathrm{SO}_n {\mathbb R}$. Using the change of variables $A' = U^T A U$ and the invariance of $g$ and of the Haar measure by conjugation, we have 
\begin{equation}
K(USU^T) = K(S).
\label{eq:invarK}
\end{equation} 
The space ${\mathcal S}_n^0$ is an irreducible representation of $\mathrm{SO}_n {\mathbb R}$ (see Lemma \ref{lem:SOnirrep} below) and ${\mathbb R}$ is the trivial representation, which is also irreducible. Formula \eqref{eq:invarK} says that the linear map $K$ intertwins these two irreducible representations. Since they are not isomorphic (they do not have the same dimension), by Schur's Lemma (see \cite[Lemma 1.7]{Fulton_Harris} or Section \ref{sec:rep_theory}), $K$ must be identically $0$. 

Specifying $g =  M_{\mathrm{Id}}$, we now prove the properties of  $c_1$ stated in the second part of Lemma \ref{lem:int_A_VM_dA}. First, we show that $c_1(0) = 0$. Indeed, 
$$ c_1(0) = \frac{1}{n} \int_{\mathrm{SO}_n{\mathbb R}} \mathrm{Tr} A \, d A. $$
The function $\mathrm{SO}_n{\mathbb R} \ni A \mapsto \mathrm{Tr} \, A \in {\mathbb R}$ is the character of the standard representation, while the function $\mathrm{SO}_n{\mathbb R} \ni A \mapsto 1 \in {\mathbb R}$ is the character of the trivial representations. Since both are irreducible, by \eqref{eq:Schur_ortho}, we get $c_1(0) = 0$. Now, we show that $c_1$ is nondecreasing. By differentiating \eqref{eq:int_A_VM_dA} with respect to $\kappa$, we get $n \frac{d c_1}{d \kappa} = \frac{N}{D}$ with 
$$ D = \Big( \int_{\mathrm{SO}_n{\mathbb R}} \mathrm{Tr} A \, e^{\kappa \mathrm{Tr} A} \, d A \Big)^2 > 0, $$
and 
\begin{eqnarray*}
N &=& \int_{(\mathrm{SO}_n{\mathbb R})^2} e^{\kappa \mathrm{Tr} A} \, e^{\kappa \mathrm{Tr} A'} \Big[ (\mathrm{Tr} \, A)^2 - \mathrm{Tr} A \, \mathrm{Tr} A' \Big] \, dA \, dA' \\
&=& \frac{1}{2} \int_{(\mathrm{SO}_n{\mathbb R})^2} e^{\kappa \mathrm{Tr} A} \, e^{\kappa \mathrm{Tr} A'} \Big[ \mathrm{Tr} A -  \mathrm{Tr} A' \Big]^2 \, dA \, dA' \geq 0. 
\end{eqnarray*}
We finally show that $c_1(\kappa) \to 1$ as $\kappa \to \infty$. This is a classical method of concentration of measure. Define the probability measures 
$$d \mu_\kappa (A) = \frac{e^{\kappa \mathrm{Tr} A} \, dA}{\int_{\mathrm{SO}_n{\mathbb R}} e^{\kappa \mathrm{Tr} A} \, dA}. $$
Since $\mathrm{SO}_n{\mathbb R}$ is compact, the family $\{\mu_\kappa\}_{\kappa \in [0,\infty)}$ is tight, so there is a sequence $(\kappa_n)_{n \geq 1}$, $\kappa_n \to \infty$ and a probability measure $\mu$ on  $\mathrm{SO}_n{\mathbb R}$ such that $\mu_{\kappa_n}$ converges weakly to $\mu$, i.e. for any measurable subset $S$ of $\mathrm{SO}_n{\mathbb R}$, $\mu_{\kappa_n} (S) \to \mu (S)$. Now, the support of $\mu$ is reduced to the singleton $\{ \mathrm{Id} \}$, so that it is the Dirac delta $\mu = \delta_{\mathrm{Id}}$. Indeed, we show that for any $B \in \mathrm{SO}_n{\mathbb R}$, $B \not = \mathrm{Id}$, there exists an open set $W$ containing $B$ such that $\mu(W) = 0$. First, we note that since $B \not = \mathrm{Id}$, then, $\mathrm{Tr} B < n = \mathrm{Tr} \, \mathrm{Id}$. This is because $B$ is conjugate to one of the matrices $R_{\theta_1, \ldots, \theta_p}$ and that $\mathrm{Tr} \, R_{\theta_1, \ldots, \theta_p} \leq n$ with equality if and only if $\cos \theta_1 = \ldots = \cos \theta_p = 1$, i.e. $R_{\theta_1, \ldots, \theta_p} = \mathrm{Id}$. Let $c$, $c'$ be two constants such that $\frac{n + \mathrm{Tr} B}{2} < c < c' < n$ and define $W = \mathrm{Tr}^{-1} \big( (- \infty, c) \big)$ and $W' = \mathrm{Tr}^{-1} \big( (c', \infty) \big)$. Then, 
\begin{eqnarray*}
\mu_\kappa(W) &=& \frac{\int_W e^{\kappa \mathrm{Tr} A} \, dA}{\int_{\mathrm{SO}_n{\mathbb R}} e^{\kappa \mathrm{Tr} A} \, dA} \leq \frac{\int_W e^{\kappa \mathrm{Tr} A} \, dA}{\int_{W'} e^{\kappa \mathrm{Tr} A} \, dA} \\
&\leq & \frac{e^{\kappa c} m(W)}{e^{\kappa c'} m(W')} \to 0 \textrm{ as } \kappa \to \infty, 
\end{eqnarray*}
where $m(W)$, $m(W')$ denote the Haar measures of $W$ and $W'$. Now, since the limit of all convergent subsequences is $\delta_{\mathrm{Id}}$, the whole family $\mu_\kappa$ converges weakly to $\delta_{\mathrm{Id}}$. In particular, since $c_1(\kappa) = \frac{1}{n} \langle \mu_{\kappa}, \mathrm{Tr} A \rangle$, this implies that 
$$ \lim_{\kappa \to \infty} c_1(\kappa) = \frac{1}{n} \langle \delta_{\mathrm{Id}}, \mathrm{Tr} A \rangle = \frac{1}{n} \mathrm{Tr} \, \mathrm{Id} = 1. $$
This ends the proof of Lemma \ref{lem:int_A_VM_dA}. \endproof

\begin{remark}
To explore the properties of  $c_1$, one could also use \eqref{eq:express_c1_WIF}. From $C_{2p}^{(1)} \leq 2p$ and $C_{2p+1}^{(1)} \leq 2p+1$ it follows that $c_1 \leq 1$. The other properties of $c_1$ are easy to prove in the case $n=2p$. For instance, to prove $c_1 \geq 0$ it is enough to show that the numerator of \eqref{eq:express_c1_WIF} is nonnegative. Then, we note that $C_{2p}^{(1)}$ is changed in its opposite when $(\theta_1, \ldots, \theta_i, \ldots, \theta_p)$ is changed into $(-\theta_1, \ldots, -\theta_i, \ldots, -\theta_p)$, while $u_{2p}$ is unchanged. Consequently, defining $D = (C_{2p}^{(1)})^{-1} \big((0,\infty) \big)$, we can write 
$$ \int_{[0,\pi]^p} C_{2p}^{(1)} \, \exp \big( \kappa C_{2p}^{(1)} \big) \, u_{2p} \, d \tilde \theta_p = 2 \int_{D} C_{2p}^{(1)} \, \sinh \big( \kappa C_{2p}^{(1)} \big) \, u_{2p} \, d \tilde \theta_p \geq 0. $$
The same method would permit to show that $\kappa \mapsto c_{2p}(\kappa)$ is non decreasing and that $c_{2p}(0) = 0$. This is entirely different in the case $n=2p+1$. For instance, that $c_{2p+1}(0)=0$ amounts to the identity
\begin{eqnarray*}
&&\hspace{-1cm} 
 \int_{[0,\pi]^p} \big( 2(\cos  \theta_1 + \ldots + \cos  \theta_p) + 1 \big) \, \exp \Big( \kappa \big( 2(\cos  \theta_1 + \ldots + \cos  \theta_p) + 1 \big) \Big) \\
&&\hspace{3cm} 
\times \prod_{1 \leq j<k \leq p} \big( \cos \theta_j - \cos \theta_k \big)^2 \, \prod_{j=1}^p \big(1 - \cos \theta_j \big) \, d \theta_1 \ldots d\theta_p = 0, 
\end{eqnarray*}
which does not look obvious to show directly. Similarly, the positivity of $c_{2p+1}(\kappa)$ or that $\kappa \mapsto c_{2p+1}(\kappa)$ is increasing are not obvious as well. 
\end{remark}

\begin{lemma}
The space ${\mathcal S}_n^0$ is an irreducible representation of $\mathrm{SO}_n {\mathbb R}$.
\label{lem:SOnirrep}
\end{lemma}

\noindent
\textbf{Proof.} This fact is classical but we sketch it here as a warm-up for the use of the concepts of Section \ref{sec:rep_theory}. The space of symmetric matrices with complex entries is isomorphic to $\textrm{Sym}^2(V)$ with $V = {\mathbb C}^n$, which is the Weyl module ${\mathbb S}_2(V)$ and thus, an irreducible representation of $\mathfrak{sl}_n{\mathbb C}$. We use Weyl's construction using the contractions (see Section \ref{sec:rep_theory} or \cite[\S 19.5]{Fulton_Harris}), to find its associated irreducible representations over $\mathfrak{so}_n{\mathbb C}$. There is only one contraction 
\begin{equation}
\Phi: \left\{ \begin{array}{lll} 
\textrm{Sym}^2(V) & \rightarrow & {\mathbb C}  \\
v_1 \circ v_2 & \to & 2 (v_1 \cdot v_2).  
\end{array} \right.  \label{eq:defPhi}
\end{equation}
The kernel of this map, 
\begin{equation} 
{\mathbb S}_{[2]}(V) = \textrm{Span} \{ v_1 \circ v_2  \, \, | \, \, v_1, \, v_2 \in V, \textrm{ such that } v_1 \cdot v_2= 0 \}. 
\label{eq:S[2]}
\end{equation}
is an irreducible representation of $\mathfrak{so}_n{\mathbb C}$. In terms of matrices, ${\mathbb S}_{[2]}(V)$ is nothing but the space of trace-free symmetric matrices with complex entries. Thus, ${\mathbb S}_{[2]}(V)$ is a complex irreducible representation of $\mathrm{SO}_n{\mathbb R}$ and is of real type (see Section \ref{sec:rep_theory}) so its real part ${\mathcal S}_n^0$ is an irreducible real representation of $\mathrm{SO}_n{\mathbb R}$. \endproof


\setcounter{equation}{0}
\section{Proof of Lemma \ref{lem:expressLB} (i)}
\label{sec:proof:expressLBi}

Again, we show that \eqref{eq:expressL} is true in the more general case where $M_{\mathrm{Id}}$ is replaced by any class function $g$ (but there is no need to suppose that $g$ is invariant by transposition). By linearity, we extend $L$ given by \eqref{eq:defL} to a mapping $\tilde L$: $\tilde {\mathcal A}_n \to \tilde {\mathcal A}_n$, where $\tilde {\mathcal A}_n$ is the complexification of ${\mathcal A}_n$, i.e. the space of antisymmetric matrices with complex entries. Thus,  $\tilde L (P+iQ) = L(P) + i L(Q)$, $\forall P, \, Q \in {\mathcal A}_n$. The space $\tilde {\mathcal A}_n$ is isomorphic to the exterior square $\Lambda^2(V)$ with $V = {\mathbb C}^n$.

\medskip
\noindent
\textbf{(i) Case $n \not = 4$.}  In this case $\Lambda^2(V)$ is an irreducible representation of $\mathfrak{so}_{n}{\mathbb C}$ \cite[Theorems 19.2 and 19.14]{Fulton_Harris}. It lifts into an irreducible representation of $\mathrm{SO}_n{\mathbb C}$ and consequently, of $\mathrm{SO}_n{\mathbb R}$. Now, $\tilde L$ is a mapping from $\Lambda^2(V)$ to itself which intertwins the two representations (i.e. $\tilde L(UPU^T) = U \tilde L(P) U^T$, $\forall U \in \mathrm{SO}_n {\mathbb R}$, by the same method as in Section \ref{sec:proof:int_A_VM_dA}). By Schur's Lemma \cite[Lemma 1.7]{Fulton_Harris}, there exists $C_2 \in {\mathbb C}$ such that $\tilde L(P+iQ) = C_2 (P+iQ)$, $\forall P, \, Q \in {\mathcal A}_n$. Taking $Q=0$, we get 
\begin{equation}
L = C_2 \, \mathrm{Id}_{\Lambda^2 (V)}, 
\label{eq:expressL_ngene}
\end{equation}
and since $L(P)$ has real entries, we have $C_2 \in {\mathbb R}$. 

We now show the second formula of \eqref{eq:expressL}. Taking the matrix inner product of $L(P) = C_2 \, P$ with $P$ leads to 
$$ \int_{\mathrm{SO}_n {\mathbb R}}  (A \cdot P)^2 \, g(A) \, dA = C_2 P\cdot P, $$
and taking $P = e_i \wedge e_j$ for $i \not = j$ gives 
$$ \int_{\mathrm{SO}_n {\mathbb R}}  (A_{ij} - A_{ji})^2 \, g(A) \, dA = 2 C_2 (\delta_{ii} \, \delta_{jj} - \delta_{ij} \, \delta_{ji}), $$
where $A = (A_{ij})_{i, j = 1, \ldots, n}$. We note that the formula is still valid for $i=j$. Summing over $i, \, j$, expanding the square, and noting that 
\begin{equation}
\sum_{i, j} A_{ij}^2 = \mathrm{Tr} (A^T A) = \mathrm{Tr} \, \mathrm{Id} = n, \quad \sum_{i, j} A_{ij} A_{ji} = \mathrm{Tr} \, A^2, 
\label{eq:traces}
\end{equation}
we arrive at the second formula of \eqref{eq:expressL} when $g =  M_{\mathrm{Id}}$. 

\medskip
\noindent
\textbf{(ii) Case $n = 4$.} $\Lambda^2(V)$ (with $V = {\mathbb C}^4$) is not an irreducible representation of $\mathfrak{so}_{4}{\mathbb C}$. It decomposes into the direct sum of two non-isomorphic irreducible representations of $\mathfrak{so}_{4}{\mathbb C}$ \cite[Theorem 19.2 (ii)]{Fulton_Harris}: 
\begin{equation} 
\Lambda^2(V) = \Lambda_+ \oplus \Lambda_-, 
\label{eq:decompL2}
\end{equation}
both having dimension $3$ (we remind that $\textrm{dim} \, \Lambda^2(V) = {4 \choose 2} = 6$). Furthermore, $\Lambda_+$ and $\Lambda_-$ lift into complex irreducible representations of $\mathrm{SO}_4{\mathbb C}$ and thus, of $\mathrm{SO}_4{\mathbb R}$ \cite[Proposition 23.13 (iii)]{Fulton_Harris}. Let ${\mathcal T}_\pm$: $\Lambda^2(V) \to \Lambda_\pm$ be the projections of $\Lambda^2(V)$ on these two sub-representations. The map $\tilde L$ can be decomposed by blocks using \eqref{eq:decompL2} on both its domain and codomain. Each block being a complex irreducible representation of $\mathrm{SO}_4{\mathbb R}$, we can apply Schur's lemma and conclude that any map between two blocks is equal to~$0$ if the blocks are not isomorphic and equal to $C \, \mathrm{Id}$ for some constant $C \in {\mathbb C}$ if the blocks are isomorphic. The pairs of isomorphic blocks are $(\Lambda_+, \Lambda_+)$ and $(\Lambda_-, \Lambda_-)$. It follows that there exist two constants $C_2^+, \, C_2^- \in {\mathbb C}$ such that 
\begin{equation} 
\tilde L  = C_2^+ \, {\mathcal T}_+ + C_2^- \, {\mathcal T}_- ,  
\label{eq:express_tilL_n4}
\end{equation}

We now compute ${\mathcal T}_\pm$. We have an automorphism of $\mathrm{SO}_4{\mathbb R}$ representations $\alpha$: $\Lambda^2 (V) \to \Lambda^2 (V)$ given by 
\begin{equation}
(\alpha(v_1 \wedge v_2) \cdot v_3 \wedge v_4) = \det (v_1, v_2, v_3, v_4), \quad  \forall (v_1, \ldots, v_4) \in V^4.
\label{eq:def_alpha_n4}
\end{equation}
We recall that the inner product on $\Lambda^2 (V)$ is given by $(v_1 \wedge v_2 \cdot v_3 \wedge v_4) = 2 [ (v_1 \cdot v_3) (v_2 \cdot v_4) - (v_1 \cdot v_4) (v_2 \cdot v_3)]$. A simple computation shows that 
$$ \alpha(e_i \wedge e_j) = \frac{1}{4} \sum_{k, \ell = 1}^4 \varepsilon_{ijk\ell} \,  e_k \wedge e_\ell, $$
where $\varepsilon_{ijk\ell} = 0$ if two or more indices among $\{i, j, k, \ell\}$ are equal, and is the signature of the permutation $1 \to i$, $2 \to j$, $3 \to k$, $4 \to \ell$ otherwise. We note that 
$$ \sum_{k, \ell = 1}^4 \varepsilon_{ijk\ell} \, \varepsilon_{k \ell mn} = 2 (\delta_{im} \, \delta_{jn} - \delta_{in}  \, \delta_{jm}). $$
It follows that $\alpha^2 (e_i \wedge e_j) = \frac{1}{4} e_i \wedge e_j$, hence $(2 \alpha)^2 = \mathrm{Id}_{\Lambda^2 (V)}$. Since clearly $2\alpha \not = \pm \mathrm{Id}_{\Lambda^2 (V)}$, the eigenvalues of $2 \alpha$ are $\pm 1$. The associated eigenspaces are sub-representations of $\Lambda^2 (V)$. Since the only sub-representations of $\Lambda^2 (V)$ are $\Lambda_{\pm}$ which are irreducible, these eigenspaces coincide with $\Lambda_{\pm}$. We let $\Lambda_+$ be the eigenspace associated with eigenvalue $1$ and $\Lambda_-$ with eigenvalue $-1$. Additionally, we see from \eqref{eq:def_alpha_n4} that $\alpha$ is self-adjoint. Thus, $\Lambda_+$ and $\Lambda_-$ are orthogonal. It follows that $2\alpha$ is the orthogonal reflection of $\Lambda^2 (V)$ in the subspace $\Lambda_+$. The projections ${\mathcal T}_\pm$ are given by ${\mathcal T}_\pm  = \frac{1}{2} \mathrm{Id}_{\Lambda^2 (V)} \pm \alpha$. From \eqref{eq:express_tilL_n4}, we deduce that 
\begin{equation} 
\tilde L  = \frac{C_2^+ + C_2^-}{2} \, \mathrm{Id}_{\Lambda^2 (V)} +  (C_2^+ - C_2^-) \alpha.  
\label{eq:express_tilL_n4_2}
\end{equation}

Now, we additionally assume that $g$ is invariant by all automorphisms of $\mathrm{SO}_4{\mathbb C}$ defined by $A \mapsto UAU^T$ when $U$ ranges in $\mathrm{O}_4{\mathbb C}$. When $U \in \mathrm{O}_4{\mathbb C} \setminus \mathrm{SO}_4{\mathbb C}$, these automorphisms consist of the conjugation by an element $U$ which does not belong to $\mathrm{SO}_4{\mathbb C}$ although the image $UAU^T$ still belongs to $\mathrm{SO}_4{\mathbb C}$, so they are outer automorphisms. We note that $g = M_{\mathrm{Id}}$ satisfies this assumption. Since the Haar measure on $\mathrm{SO}_4{\mathbb C}$ is the restriction to $\mathrm{SO}_4{\mathbb C}$ of the Haar measure of $\mathrm{O}_4{\mathbb C}$ (up to a normalization factor), it is invariant by these outer automorphisms. It follows that $\tilde L$ satisfies $\tilde L(UPU^T) = U \tilde L(P) U^T$, $\forall U \in \textrm{O}_4 {\mathbb R}$, i.e., $\tilde L$ is an intertwining map for $\mathrm{O}_4{\mathbb C}$, not only $\mathrm{SO}_4{\mathbb C}$. On the other hand, $\alpha$ is alternating by outer-automorphisms, i.e. it satisfies $\alpha (UPU^T) = \det U \, U \alpha(P) U^T$. This is a consequence of \eqref{eq:def_alpha_n4} and of the fact that $\det (Uv_1, \ldots, Uv_4) = \det U \, \det (v_1, \ldots, v_4)$. Taking the conjugation of \eqref{eq:express_tilL_n4_2} by $U \in \mathrm{O}_4{\mathbb C} \setminus \mathrm{SO}_4{\mathbb C}$, we get $\tilde L  = \frac{C_2^+ + C_2^-}{2} \, \mathrm{Id}_{\Lambda^2 (V)} -  (C_2^+ - C_2^-) \alpha$. Hence, \eqref{eq:expressL_ngene} follows with $C_2 = C_2^+ = C_2^-$ and the proof can be completed like in Case (i). \endproof


\setcounter{equation}{0}
\section{Proof of Lemma \ref{lem:expressLB} (ii)}
\label{sec:proof:expressLBii}

Like in the previous section, we show \eqref{eq:expressB} for any class function $g$ replacing $M_{\mathrm{Id}}$. By the same method as in Section \ref{sec:proof:int_A_VM_dA}, the symmetric bilinear map $B$: ${\mathcal A}_n \times {\mathcal A}_n \to {\mathcal S}_n$ satisfies 
\begin{equation}
B (U P_1 U^T, U P_2 U^T) = U B(P_1, P_2) U^T, \quad \forall P_1, P_2 \in {\mathcal A}_n, \quad \forall U \in \mathrm{SO}_n {\mathbb R}.
\label{eq:invarB}
\end{equation} 
It can be extended by linearity to the complexifications of ${\mathcal A}_n$ and ${\mathcal S}_n$. These are respectively $\Lambda^2(V)$ and $\textrm{Sym}^2 (V)$ for $V = {\mathbb C}^n$. The extended symmetric bilinear map, denoted by $\breve B$: $\Lambda^2(V) \times \Lambda^2(V) \to \textrm{Sym}^2 (V)$, is given by 
$$ \breve B (P_1 + i Q_1, P_2 + i Q_2) = B(P_1,P_2) - B(Q_1, Q_2) + i ( B(P_1,Q_2) + B(Q_1,P_2)),
$$ 
for all $P_1, \, P_2, \, Q_1, \, Q_2 \in {\mathcal A}_n$. The extended map $\breve B$ still satisfies the invariance property \eqref{eq:invarB}, now with antisymmetric matrices $P_1$, $P_2$ with complex entries. 
Due to the universal property of the symmetric product \cite[Appendix B]{Fulton_Harris}, $\breve B$ determines a unique linear map $\tilde B$: $\textrm{Sym}^2 (\Lambda^2(V)) \to \textrm{Sym}^2 (V)$ given by  
$$\tilde B \big( (v_1 \wedge v_2) \circ (w_1 \wedge w_2) \big) = \breve B(v_1 \wedge v_2, w_1 \wedge w_2), \quad \forall v_1, \, v_2, \, w_1, \, w_2 \in V. $$ 
Both $\textrm{Sym}^2 (\Lambda^2(V))$ and $\textrm{Sym}^2 (V)$ are complex representations of $\mathrm{SO}_n {\mathbb R}$. Furthermore, Eq. \eqref{eq:invarB} implies that $\tilde B$ intertwins the two representations. So, we are led to find the decompositions of $\textrm{Sym}^2 (\Lambda^2(V))$ and $\textrm{Sym}^2 (V)$ into irreducible representations of $\mathrm{SO}_n {\mathbb R}$. 

The decomposition of $\textrm{Sym}^2 (V) = {\mathbb S}_2(V)$ into irreducible representations of $\mathfrak{so}_{n}{\mathbb C}$ has been initiated in the proof of Lemma \ref{lem:SOnirrep}. What is missing is to find the supplementary representation(s) of ${\mathbb S}_{[2]}(V)$ (given by \eqref{eq:S[2]}) in $\textrm{Sym}^2 (V)$. Using \cite[p. 263]{Fulton_Harris}, we have:
\begin{eqnarray} 
\textrm{Sym}^2 (V) = {\mathbb S}_{[2]}(V) \oplus {\mathbb C} \Psi, \quad \Psi =: \sum_{i=1}^n e_i \circ e_i, 
\label{eq:decompS2}
\end{eqnarray}
The projections ${\mathcal P}_0$ and ${\mathcal P}_1$ of ${\mathbb S}_2(V)$ on ${\mathbb C} \Psi$ and ${\mathbb S}_{[2]}(V)$ are respectively given by 
\begin{equation} 
{\mathcal P}_0 (v \circ w) = \frac{1}{n} (v \cdot w) \Psi, \quad {\mathcal P}_1 = \mathrm{Id}_{{\mathbb S}_2(V)} - {\mathcal P}_0. 
\label{eq:decompS2proj}
\end{equation}
Indeed, using that $\Phi(\Psi) = 2n$, we verify that $\Phi \circ {\mathcal P}_1 = 0$ (where $\Phi$ is the contraction \eqref{eq:defPhi}), showing that $\textrm{Im} \, {\mathcal P}_1 \subset {\mathbb S}_{[2]}(V)$. In terms of matrices, $\Psi = 2 \, \mathrm{Id}$ and \eqref{eq:decompS2proj} corresponds to the decomposition
$$ S = S_0 + S_1, \quad S_0 = S - \frac{1}{n} \mathrm{Tr} \, S \, \mathrm{Id}, \quad S_1 = \frac{1}{n} \mathrm{Tr} \, S \, \mathrm{Id}, $$
of a complex symmetric matrix $S$ into a trace-free symmetric matrix $S_0$ and a scalar matrix $S_1$. Of course,  ${\mathbb S}_{[2]}(V)$ and ${\mathbb C} \Psi$ lift into complex irreducible representations of $\mathrm{SO}_n {\mathbb C}$ and thus of $\mathrm{SO}_n {\mathbb R}$ 

Now, we consider $\textrm{Sym}^2 (\Lambda^2(V))$ and first decompose it into irreducible representations of $\mathfrak{sl}_{n}{\mathbb C}$. We apply Pieri's formula \cite[Exercise 6.16]{Fulton_Harris} and decompose 
\begin{equation}
\textrm{Sym}^2 (\Lambda^2(V)) = \Lambda^4(V) \oplus {\mathbb S}_{(2,2)}(V), 
\label{eq:Sym_decomp}
\end{equation}
Being Weyl modules, both are irreducible representations of $\mathfrak{sl}_{n}{\mathbb C}$. We have the following obvious formulas for the dimensions of $\textrm{Sym}^2 (\Lambda^2(V))$ and $\Lambda^4(V)$: 
\begin{eqnarray*} 
\textrm{dim} \, \textrm{Sym}^2 (\Lambda^2(V)) &=& \frac{1}{2} \frac{n (n-1)}{2} \Big( \frac{n (n-1)}{2}  + 1 \Big) = \frac{1}{8} (n^4 - 2 n^3 + 3 n^2 - 2n), \\
\textrm{dim} \, \Lambda^4(V) &=& {n \choose 4} = \frac{1}{24} (n^4 - 6 n^3 + 11 n^2 - 6n). 
\end{eqnarray*}
On the other hand, using \cite[Theorem 6.3(i)]{Fulton_Harris}, we have 
\begin{equation} 
\textrm{dim} \, {\mathbb S}_{(2,2)}(V) = \frac{1}{12} n^2 (n^2 - 1), 
\label{eq:dimS22}
\end{equation}
and we verify that these formulas are consistent with \eqref{eq:Sym_decomp}.

The space ${\mathbb S}_{(2,2)}(V)$ in \eqref{eq:Sym_decomp} can be characterized as 
\begin{equation} 
{\mathbb S}_{(2,2)}(V)  = \textrm{Span} \{ (v_1 \wedge v_2) \circ (v_1 \wedge v_3) \, \, | \, \, v_1, \, v_2, \, v_3 \in V \}. 
\label{eq:S22}
\end{equation}
Indeed, we have a linear map $F$ of $\mathfrak{sl}_{n}{\mathbb C}$-representations 
$$ F: \left\{ \begin{array}{ccc} 
\textrm{Sym}^2 ( \Lambda^2 (V) ) & \rightarrow & \Lambda^4 (V)  \\
(v_1 \wedge v_2) \circ (v_3 \wedge v_4) & \mapsto & v_1 \wedge v_2 \wedge v_3 \wedge v_4, 
\end{array} \right. 
$$
which is clearly surjective. Its kernel $\textrm{Ker} \, F$ is not $\{0\}$ (otherwise, we would have $\textrm{Sym}^2 (\Lambda^2 (V))$ $\approx \Lambda^4 (V)$ and this cannot be the case because the dimensions are different) and is a sub-representation of $\textrm{Sym}^2 ( \Lambda^2 (V) )$. Thus, $\textrm{Ker} \, F = {\mathbb S}_{(2,2)}(V)$. Clearly $F\big((v_1 \wedge v_2) \circ (v_1 \wedge v_3)\big) = 0$, $\forall v_1, \, v_2, \, v_3 \in V$. So, $\Sigma \subset {\mathbb S}_{(2,2)}(V)$, where $\Sigma$ denotes the right-hand side of \eqref{eq:S22}. The space $\Sigma$ is a sub-representation of $\textrm{Sym}^2 ( \Lambda^2 (V) )$. Finally, $\Sigma \not = \{0\}$ (take two independent vectors $v_1$ and $v_2$. Then, $(v_1 \wedge v_2) \circ (v_1 \wedge v_2)$ belongs to $\Sigma$ and is not $0$). Thus, we must have $\Sigma = {\mathbb S}_{(2,2)}(V)$. 

Next, we find how an element of $\textrm{Sym}^2 (\Lambda^2(V))$ decomposes along \eqref{eq:Sym_decomp}. We define two endomorphisms ${\mathcal T}_1$ and ${\mathcal T}_2$ of $\textrm{Sym}^2 (\Lambda^2(V))$ which commute with the action of $\mathfrak{sl}_{n}{\mathbb C}$. They are defined by their action on a generator $(v_1 \wedge v_2) \circ (v_3 \wedge v_4)$ according to
\begin{eqnarray}
&& \hspace{-1cm}
{\mathcal T}_1 \big( (v_1 \wedge v_2) \circ (v_3 \wedge v_4) \big) = \frac{1}{3} \big[ (v_1 \wedge v_2) \circ (v_3 \wedge v_4) - (v_1 \wedge v_3) \circ (v_2 \wedge v_4) \nonumber \\
&& \hspace{2cm}
 - (v_1 \wedge v_4) \circ (v_3 \wedge v_2) \big] = \frac{1}{3} \, v_1 \wedge v_2 \wedge v_3 \wedge v_4, \label{eq:tens_decomp2}\\
&& \hspace{-1cm}
{\mathcal T}_2 \big( (v_1 \wedge v_2) \circ (v_3 \wedge v_4) \big) = \frac{1}{3} \big[ (v_1 \wedge v_2) \circ (v_3 \wedge v_4) + (v_1 \wedge v_3) \circ (v_2 \wedge v_4) \big] \nonumber \\ 
&& \hspace{2cm}
+ \frac{1}{3} \big[ (v_1 \wedge v_2) \circ (v_3 \wedge v_4) + (v_1 \wedge v_4) \circ (v_3 \wedge v_2) \big]. \label{eq:tens_decomp3}
\end{eqnarray}
We verify that 
\begin{equation}
{\mathcal T}_1 + {\mathcal T}_2 = \mathrm{Id}_{\textrm{Sym}^2 (\Lambda^2(V))}, 
\label{eq:T1+T2=Id}
\end{equation}
and that 
\begin{equation} 
\textrm{Im} \, {\mathcal T}_1  = \Lambda^4 (V), \quad \textrm{Im} \, {\mathcal T}_2  = {\mathbb S}_{(2,2)}(V). 
\label{eq:ImT1T2}
\end{equation}
Indeed, this is clear for $\textrm{Im} \, {\mathcal T}_1$. By expanding $((v_2+v_3) \wedge v_1) \circ ((v_2 + v_3) \wedge v_4) \in {\mathbb S}_{(2,2)}(V)$, we get that the first bracket at the right-hand side of \eqref{eq:tens_decomp3} is in ${\mathbb S}_{(2,2)}(V)$. We perform similarly for the second bracket with $((v_2+v_4) \wedge v_1) \circ ((v_2 + v_4) \wedge v_3)$, so that $\textrm{Im} \, {\mathcal T}_2  \subset {\mathbb S}_{(2,2)}(V)$. Because of \eqref{eq:T1+T2=Id} and \eqref{eq:Sym_decomp}, this inclusion is an equality. Therefore, ${\mathcal T}_1$ and ${\mathcal T}_2$ are the projections of $\textrm{Sym}^2 (\Lambda^2(V))$ on $\Lambda^4(V)$ and ${\mathbb S}_{(2,2)}(V)$ respectively.  

Now, $\Lambda^4(V)$ is an irreducible representation of $\mathfrak{so}_{n}{\mathbb C}$ for $n \geq 9$ \cite[Theorems 19.2 and 19.14]{Fulton_Harris}. From now on, we assume $n \geq 9$ and defer the examination of the special cases $n \in \{3, \ldots, 8 \}$ to the end of the proof. 

\medskip
\noindent
\textbf{(i) Case $n \geq 9$.} By contrast to $\Lambda^4(V)$, ${\mathbb S}_{(2,2)}(V)$ is not an irreducible representation of $\mathfrak{so}_{n}{\mathbb C}$ and we apply Weyl's construction using the contractions (see Section \ref{sec:rep_theory} or \cite[\S 19.5]{Fulton_Harris}) to decompose it in irreducible representations. There are six pairs of indices for a rank-4 tensor. For an element of ${\mathbb S}_{(2,2)}(V) \subset \textrm{Sym}^2 (\Lambda^2(V))$, contractions with respect to pairs $(1,2)$ and $(3,4)$  obviously lead to $0$. The remaining four contractions all give rise, up to a sign, to the same linear map $G$ of $\mathfrak{so}_{n}{\mathbb C}$-representations: 
$$ G: \left\{ \begin{array}{lll} 
{\mathbb S}_{(2,2)}(V) & \rightarrow & \textrm{Sym}^2(V)  \\
(z \wedge v) \circ (z \wedge w) & \mapsto & (z \cdot z) \, v \circ w + (v \cdot w) \, z \circ z 
- (z \cdot w) \, z \circ v - (z \cdot v) \, z \circ w .  
\end{array} \right. 
$$
By Weyl's construction, ${\mathbb S}_{[2,2]}(V) = \textrm{Ker} \, G$ is an irreducible representation of $\mathfrak{so}_{n}{\mathbb C}$. The map $G$ is surjective. Indeed, we check that
$$G\big( (e_i \wedge e_j) \circ (e_i \wedge e_k) \big) = e_j \circ e_k + \delta_{jk} \, e_i \circ e_i, $$
for all $i, \, j, \, k$ generate $\textrm{Sym}^2(V)$. With \eqref{eq:decompS2}, this allows us to write 
\begin{equation}
 {\mathbb S}_{(2,2)}(V)  \approx {\mathbb S}_{[2,2]}(V) \oplus \textrm{Sym}^2(V) \approx {\mathbb S}_{[2,2]}(V) \oplus {\mathbb S}_{[2]}(V) \oplus {\mathbb C}. 
\label{eq:decomposS22_0}
\end{equation}

We now need to identify the sub-representations $\Sigma_0$ and $\Sigma_1$ of ${\mathbb S}_{(2,2)}(V)$ which are isomorphic to  ${\mathbb C}$  and ${\mathbb S}_{[2]}(V)$ respectively and write how a generator of ${\mathbb S}_{(2,2)}(V)$ decomposes along \eqref{eq:decomposS22_0}. We define two linear endomorphisms ${\mathcal W}_0$ and ${\mathcal W}_1$ of ${\mathbb S}_{(2,2)}(V)$ by
\begin{equation} 
{\mathcal W}_0 \big( (z \wedge v) \circ (z \wedge w) \big) = \frac{2}{n (n-1)} \big[ (z \cdot z)  (v \cdot w) - (z \cdot v)  (z \cdot w) \big] \, \Xi,  
\label{eq:defT0}
\end{equation}
with 
$$ \Xi = \sum_{i<j} (e_i \wedge e_j) \circ (e_i \wedge e_j) = \frac{1}{2} \sum_{i, \, j} (e_i \wedge e_j) \circ (e_i \wedge e_j), $$
and 
\begin{eqnarray} 
&& \hspace{-1cm} 
{\mathcal W}_1 \big( (z \wedge v) \circ (z \wedge w) \big) = \frac{1}{n-2} \Big\{ (z \cdot z) \Big( \sum_{i=1}^n (e_i \wedge v) \circ (e_i \wedge w) - \frac{2}{n} (v \cdot w) \, \Xi \Big) \nonumber \\
&& \hspace{4.4cm}  + (v \cdot w) \Big( \sum_{i=1}^n (e_i \wedge z) \circ (e_i \wedge z) - \frac{2}{n} (z \cdot z) \, \Xi \Big) \nonumber \\
&& \hspace{4.4cm} - (z \cdot v) \Big(  \sum_{i=1}^n (e_i \wedge z) \circ (e_i \wedge w) - \frac{2}{n} (z \cdot w) \, \Xi \Big) \nonumber \\
&& \hspace{4.4cm} 
- (z \cdot w) \Big( \sum_{i=1}^n (e_i \wedge z) \circ (e_i \wedge v) - \frac{2}{n} (z \cdot v) \, \Xi \Big) \Big\}.
\label{eq:defT1}
\end{eqnarray}
These endomorphisms commute with the action of $\mathfrak{so}_{n}{\mathbb C}$. Indeed, it is at the heart of Weyl's construction to remark that the operation consisting of the insertion of $\Psi$ (see \eqref{eq:decompS2proj}) at any pair of positions inside a tensor of rank $d-2$ leading to a tensor of rank $d$ commutes with the action of $\mathfrak{so}_{n}{\mathbb C}$. The tensors $\sum_i (e_i \wedge v) \circ (e_i \wedge w)$ or $\Psi$ are obtained in this way. 

Now, We check that $ G(\Xi) = (n-1) \Psi$ and  
$$ G \big( \sum_{i=1}^n (e_i \wedge v) \circ (e_i \wedge w) \big) = (n-2) v \circ w + (v \cdot w) \Psi, \quad \forall v, \, w \in V, $$
so that:
\begin{equation} 
(G \circ {\mathcal W}_0) \big( (z \wedge v) \circ (z \wedge w) \big)  = \frac{2}{n} \big[ (z \cdot z)  (v \cdot w) - (z \cdot v)  (z \cdot w) \big] \, \Psi, 
\label{eq:GoW0}
\end{equation}
and 
\begin{equation} 
G \circ ({\mathcal W}_0 + {\mathcal W}_1) = G. 
\label{eq:Go(W0+W1)=G}
\end{equation} 
It follows that 
\begin{equation} 
\textrm{Im} \Big( \mathrm{Id}_{{\mathbb S}_{(2,2)}(V)} - {\mathcal W}_0 - {\mathcal W}_1 \Big) \subset {\mathbb S}_{[2,2]}(V). 
\label{eq:ImId-W0-W1}
\end{equation}
Now, $\textrm{Im} \, G \circ {\mathcal W}_0$ and $\textrm{Im} \, G \circ {\mathcal W}_1$ are sub-representations of $\textrm{Sym}^2(V)$ and, owing to the fact that the bracket at the right-hand side of \eqref{eq:GoW0} can be non-zero, $\textrm{Im} \, G \circ {\mathcal W}_0 = {\mathbb C} \Psi$. Now, $\textrm{Im} \, G \circ {\mathcal W}_1 \not = \{0\}$. Otherwise, from \eqref{eq:Go(W0+W1)=G}, $\textrm{Im} \, G = \textrm{Im} \, G \circ {\mathcal W}_0 = {\mathbb C} \Psi$ which contradicts the fact that $G$ is surjective on $\textrm{Sym}^2(V)$. Let $\Phi$ be the contraction \eqref{eq:defPhi}. We have $\Phi(\Psi) = 2n$, from which we deduce that 
$$ (\Phi \circ G \circ {\mathcal W}_0) \big( (z \wedge v) \circ (z \wedge w) \big) = 4 \big[ (z \cdot z)  (v \cdot w) - (z \cdot v)  (z \cdot w) \big] , $$
and thus 
\begin{equation} 
\Phi \circ G \circ {\mathcal W}_0 = \Phi \circ G, \qquad \Phi \circ G \circ {\mathcal W}_1 = 0.  
\label{eq:PhioGoW}
\end{equation}
So, $\textrm{Im} \, G \circ {\mathcal W}_1$ is a sub-representation of ${\mathbb S}_{[2]}(V)$. But ${\mathbb S}_{[2]}(V)$ is irreducible and $\textrm{Im} \, G \circ {\mathcal W}_1 \not = \{0\}$, thus $\textrm{Im} \, G \circ {\mathcal W}_1 = {\mathbb S}_{[2]}(V)$. 

We define the following sub-representations of ${\mathbb S}_{(2,2)}(V)$:
$$ \Sigma_0 = \textrm{Im} \, {\mathcal W}_0, \qquad \Sigma_1 = \textrm{Im} \, {\mathcal W}_1. $$
From \eqref{eq:defT0}, \eqref{eq:defT1}, we have: 
\begin{eqnarray}
\Sigma_0 &=&  {\mathbb C} \, \Xi, \nonumber \\
\Sigma_1 &=& \textrm{Span} \Big\{\sum_{i=1}^n (e_i \wedge v) \circ (e_i \wedge w) - \frac{2}{n} (v \cdot w) \Xi \quad \big| \quad v, \, w \in V \Big\}.  \label{eq:express_Sig1}
\end{eqnarray}
From \eqref{eq:ImId-W0-W1}, we have 
\begin{equation} 
{\mathbb S}_{(2,2)}(V) = \Sigma_0 + \Sigma_1 + {\mathbb S}_{[2,2]}(V). 
\label{eq:sum}
\end{equation} 
All the spaces at the right-hand side of \eqref{eq:sum} are sub-representations of ${\mathbb S}_{(2,2)}(V)$ and we know that $\Sigma_0$ and ${\mathbb S}_{[2,2]}(V)$ are irreducible. Furthermore, according to \cite[Formulas 24.29 and 24.41]{Fulton_Harris}, we have 
\begin{equation}
\textrm{dim} \, {\mathbb S}_{[2,2]}(V) = \frac{1}{12} (n^4 - 7 n^2 - 6n). 
\label{eq:dimS[22]}
\end{equation}
We deduce that $\Sigma_0$ and ${\mathbb S}_{[2,2]}(V)$ are not isomorphic, since $\textrm{dim} \,\Sigma_0 = 1$ is different from $\textrm{dim} \, {\mathbb S}_{[2,2]}(V)$. This implies that $\Sigma_0 \cap {\mathbb S}_{[2,2]}(V) = \{0\}$. Suppose now $\Sigma_1 \cap {\mathbb S}_{[2,2]}(V) \not = \{0\}$. Then, $\Sigma_1 \cap {\mathbb S}_{[2,2]}(V) = {\mathbb S}_{[2,2]}(V)$, i.e. ${\mathbb S}_{[2,2]}(V) \subset \Sigma_1$. But $\textrm{dim} \, \Sigma_1 \leq \frac{n(n+1)}{2}$, while, $\textrm{dim} \, {\mathbb S}_{[2,2]}(V)$ is given by \eqref{eq:dimS[22]}. We easily check that for $n \geq 5$, we cannot have $\textrm{dim} \, {\mathbb S}_{[2,2]}(V) \leq \frac{n(n+1)}{2}$ (the special cases of small dimension will be examined later). So, we must have $\Sigma_1 \cap {\mathbb S}_{[2,2]}(V) = \{0\}$. Finally, assume that $\lambda \Xi \in \Sigma_1$. Thanks to \eqref{eq:PhioGoW}, we have $(\Phi \circ G) (\lambda \Xi)= 0$ on the one hand and $(\Phi \circ G) (\lambda \Xi) = 2 n (n-1) \lambda$. This implies $\lambda = 0$ and consequently, $\Sigma_0 \cap \Sigma_1 = \{0\}$. It follows that the sum \eqref{eq:sum} is a direct sum and that we have 
\begin{equation} 
{\mathbb S}_{(2,2)}(V) = \Sigma_0 \oplus \Sigma_1 \oplus {\mathbb S}_{[2,2]}(V).  
\label{eq:decompS22}
\end{equation}
Additionally, in view of \eqref{eq:decomposS22_0}, $\Sigma_1 \approx {\mathbb S}_{[2]}(V)$. In view of \eqref{eq:ImId-W0-W1}, ${\mathcal W}_0$, ${\mathcal W}_1$ and $1 - ({\mathcal W}_0 + {\mathcal W}_1)$ are the projections associated to this direct sum, respectively on the spaces $\Sigma_0$, $\Sigma_1$ and ${\mathbb S}_{[2,2]}(V)$. One last remark is that the description of $\Sigma_1$ can be slightly simplified from \eqref{eq:express_Sig1}. We have 
\begin{equation}
\Sigma_1 = \textrm{Span} \Big\{\sum_{i=1}^n (e_i \wedge v) \circ (e_i \wedge w) \quad \big| \quad v, \, w \in V \textrm{ such that } v \cdot w = 0 \Big\}. 
\label{eq:express_Sig1_alter}
\end{equation}
Indeed, the right-hand side of \eqref{eq:express_Sig1_alter} is a sub-representation of $\Sigma_1$ which is obviously not $\{0\}$ and since $\Sigma_1$ is irreducible, the two spaces must be equal. With \eqref{eq:express_Sig1_alter}, the isomorphism ${\mathcal J}_1$ between ${\mathbb S}_{[2]}(V)$ and $\Sigma_1$ is simply given by 
\begin{equation} 
{\mathcal J}_1 (v \circ w) = \sum_{i=1}^n (e_i \wedge v) \circ (e_i \wedge w), \quad \forall v, \, w \in V \textrm{ such that } v \cdot w = 0. 
\label{eq:expressJ1}
\end{equation}
Obviously, the isomorphism ${\mathcal J}_0$ between ${\mathbb C} \Psi$ and $\Sigma_0$ is given by 
\begin{equation} 
{\mathcal J}_0 (\lambda \Psi) = \lambda \Xi, \quad \forall \lambda \in {\mathbb C}. 
\label{eq:expressJ0}
\end{equation}
Finally, with $\textrm{dim} \, \Sigma_0 = 1$ and $\textrm{dim} \, \Sigma_1 = \frac{n(n+1)}{2} - 1$, \eqref{eq:dimS[22]} and \eqref{eq:dimS22}, we check the consistency of the dimensions with \eqref{eq:decompS22}.

Collecting \eqref{eq:Sym_decomp} and \eqref{eq:decompS22}, we get the decomposition of $\textrm{Sym}^2 (\Lambda^2(V))$: 
\begin{equation}
\textrm{Sym}^2 (\Lambda^2(V)) = \Sigma_0 \oplus \Sigma_1 \oplus {\mathbb S}_{[2,2]}(V) \oplus \Lambda^4(V), 
\label{eq:decompS2(Lamb)}
\end{equation}
and we denote by ${\mathcal Q}_0\, \ldots, {\mathcal Q}_3$, the projections of $\textrm{Sym}^2 (\Lambda^2(V))$ on $\Sigma_0$, $\Sigma_1$, ${\mathbb S}_{[2,2]}(V)$ and $\Lambda^4(V)$ respectively. In particular, we have 
\begin{equation} 
{\mathcal Q}_0 = {\mathcal W}_0 \circ {\mathcal T}_2, \quad {\mathcal Q}_1 = {\mathcal W}_1 \circ {\mathcal T}_2, 
\label{eq:defQ0Q1}
\end{equation}
where we have restricted the codomain of ${\mathcal T}_2$ to ${\mathbb S}_{(2,2)}(V)$. In the case $n \geq 9$, \eqref{eq:decompS2(Lamb)} is the decomposition of $\textrm{Sym}^2 (\Lambda^2(V))$ into irreducible representations of $\mathfrak{so}_n{\mathbb C}$. Since all these representations are of the form ${\mathbb S}_{[\lambda]}(V)$ for an appropriate partition $\lambda$ of $4$, they all lift into complex irreducible representations of $\mathrm{SO}_n{\mathbb C}$ and thus, of $\mathrm{SO}_n{\mathbb R}$. 

Now, we express how ${\mathcal Q}_0$ and ${\mathcal Q}_1$ act on a generator of $\textrm{Sym}^2 (\Lambda^2(V))$. Inserting \eqref{eq:tens_decomp3}, \eqref{eq:defT0} and \eqref{eq:defT1} into \eqref{eq:defQ0Q1}, we find: 
\begin{equation} 
{\mathcal Q}_0 \big( (v_1 \wedge v_2) \circ (v_3 \wedge v_4) \big) = \frac{2}{n (n-1)} \big[ (v_1 \cdot v_3)  (v_2 \cdot v_4) - (v_1 \cdot v_4)  (v_2 \cdot v_3) \big] \, \Xi,  
\label{eq:defQ0}
\end{equation}
and 
\begin{eqnarray} 
&& \hspace{-1cm} 
{\mathcal Q}_1 \big( (v_1 \wedge v_2) \circ (v_3 \wedge v_4) \big) = \frac{1}{n-2} \Big\{ (v_1 \cdot v_3) \Big( \sum_{i=1}^n (e_i \wedge v_2) \circ (e_i \wedge v_4) - \frac{2}{n} (v_2 \cdot v_4) \, \Xi \Big) \nonumber \\
&& \hspace{3.4cm}  + (v_2 \cdot v_4) \Big( \sum_{i=1}^n (e_i \wedge v_1) \circ (e_i \wedge v_3) - \frac{2}{n} (v_1 \cdot v_3) \, \Xi \Big) \nonumber \\
&& \hspace{3.4cm} - (v_1 \cdot v_4) \Big(  \sum_{i=1}^n (e_i \wedge v_2) \circ (e_i \wedge v_3) - \frac{2}{n} (v_2 \cdot v_3) \, \Xi \Big) \nonumber \\
&& \hspace{3.4cm} 
- (v_2 \cdot v_3) \Big( \sum_{i=1}^n (e_i \wedge v_1) \circ (e_i \wedge v_4) - \frac{2}{n} (v_1 \cdot v_4) \, \Xi \Big) \Big\}.
\label{eq:defQ1}
\end{eqnarray}
Then, let $P, \, Q \in \Lambda^2(V)$ and write their decomposition in the basis $\{e_i \wedge e_j\}_{i<j}$ as follows: 
$$ P = \sum_{1 \leq i<j \leq n} P_{ij} \, e_i \wedge e_j = \frac{1}{2} \sum_{i,j=1}^n P_{ij} \, e_i \wedge e_j, $$
where $P_{ji} = - P_{ij}$ for $i>j$ and $P_{ii} = 0$, and similarly for $Q$. We identify $P$ and $Q$ with the matrices $(P_{ij})_{i,j=1}^n$ and $(Q_{ij})_{i,j=1}^n$. Application of \eqref{eq:defQ0} and \eqref{eq:defQ1} lead to 
\begin{equation} 
{\mathcal Q}_0 \big( P \circ Q \big) = - \frac{1}{n (n-1)} \, \mathrm{Tr} (PQ) \, \Xi,  
\label{eq:Q0(PQ)}
\end{equation}
and 
\begin{eqnarray} 
&&\hspace{-1cm}
{\mathcal Q}_1 \big( P \circ Q \big) = - \frac{1}{n-2} \, \Big\{ \sum_{i,j = 1}^n (PQ)_{ij} \sum_{m=1}^n (e_m \wedge e_i) \circ (e_m \wedge e_j) - \frac{2}{n} \, \mathrm{Tr} (PQ) \, \Xi \Big\} \nonumber \\
&& \hspace{0cm} = - \frac{1}{n-2}  \sum_{i,j = 1}^n \Big( \frac{PQ+QP}{2} - \frac{1}{n} \mathrm{Tr} (PQ) \, \mathrm{Id} \Big)_{ij} \sum_{m=1}^n (e_m \wedge e_i) \circ (e_m \wedge e_j), \label{eq:Q1(PQ)}
\end{eqnarray}
with $PQ$ is the matrix product of $P$ and $Q$, $\mathrm{Tr} (PQ)$, its trace and $(PQ)_{ij}$, its $(i,j)$-th entry. 

Now, the map $\tilde B$ can be decomposed by blocks using \eqref{eq:decompS2(Lamb)} on the domain and \eqref{eq:decompS2} on the codomain. Applying the same arguments as in Section \ref{sec:proof:expressLBi}, Case (ii), there exist two constants $C'_3, \, C'_4 \in {\mathbb C}$ such that 
\begin{equation} 
\tilde B  = C'_3 \, {\mathcal J}_0^{-1} \circ {\mathcal Q}_0 + C'_4 \, {\mathcal J}_1^{-1} \circ {\mathcal Q}_1,  
\label{eq:express_tilB}
\end{equation}
i.e., inserting \eqref{eq:expressJ1}, \eqref{eq:expressJ0}, \eqref{eq:Q0(PQ)}, \eqref{eq:Q1(PQ)} into \eqref{eq:express_tilB}, there exist two constants $C_3, \, C_4 \in {\mathbb C}$ such that \eqref{eq:expressB} holds with $B$ replaced by $\breve B$ and  any $P, Q \in \Lambda^2(V)$. Now, if we take $P$ and $Q$ real, i.e. belonging to ${\mathcal A}_n$, then, $\breve B(P,Q)=B(P,Q)$ is real which implies that $C_3, \, C_4 \in {\mathbb R}$ and finishes the proof of  \eqref{eq:expressB}.

We now show \eqref{eq:expressc3} and \eqref{eq:expressc4}. We have
\begin{eqnarray}
{\mathcal P}_0 \circ B(P,Q) &=& \int_{\mathrm{SO}_n {\mathbb R}} (A \cdot P) \, (A \cdot Q) \, \frac{1}{n} \, \mathrm{Tr} A \, g(A) \, dA \, \mathrm{Id}, \label{eq:epressP0oB} \\
{\mathcal P}_1 \circ B(P,Q) &=& \int_{\mathrm{SO}_n {\mathbb R}} (A \cdot P) \, (A \cdot Q) \, \Big( \frac{A + A^T}{2} - \frac{1}{n} \, \mathrm{Tr} A \, \mathrm{Id} \Big) \, g(A) \, dA. \label{eq:epressP1oB}
\end{eqnarray}
Comparing \eqref{eq:epressP0oB} with \eqref{eq:expressB}, we get 
$$ C_3 \, \mathrm{Tr} (PQ) =  \frac{1}{n} \, \int_{\mathrm{SO}_n {\mathbb R}} (A \cdot P) \, (A \cdot Q) \,\mathrm{Tr} A \, g(A) \, dA, \quad \forall P, \, Q \in {\mathcal A}_n. $$
Taking 
\begin{equation}
P = e_i \wedge e_j, \, \, Q = e_k \wedge e_\ell, \quad \forall i, \, j, \, k, \, \ell \in \{1, \ldots, n \}, \quad i \not = j \textrm{ and } k \not = \ell, 
\label{eq:choicePiQi}
\end{equation}
we get 
$$ 2 C_3 \, (\delta_{jk} \, \delta_{i \ell} - \delta_{j \ell} \delta_{ik}) = \frac{1}{n} \, \int_{\mathrm{SO}_n {\mathbb R}} (A_{ij}-A_{ji}) (A_{k\ell} - A_{\ell k}) \, \mathrm{Tr} A \, g(A) \, d A. $$
we note that both sides are zero if $i=j$ or $k = \ell$ so the equality is valid for any $i, \, j, \, k, \, \ell \in \{1, \ldots, n \}$. Taking $k=j$ and $\ell = i$, and summing over all $i, \, j \in \{1, \ldots, n \}$, we get 
\begin{equation} 
C_3 = - \frac{1}{2n(n^2 - n)} \, \int_{\mathrm{SO}_n {\mathbb R}} \sum_{i, j=1}^n (A_{ij}-A_{ji})^2 \, \mathrm{Tr} A \, g(A) \, d A.
\label{eq:calculc3}
\end{equation}
inserting \eqref{eq:traces} into \eqref{eq:calculc3} and making $g = M_{\mathrm{Id}}$ leads to \eqref{eq:expressc3}. 

Now, comparing \eqref{eq:epressP1oB} with \eqref{eq:expressB}, we get 
\begin{eqnarray}
&&\hspace{-1cm} C_4 \, \Big( \frac{PQ+QP}{2} - \frac{1}{n} \mathrm{Tr} (PQ) \,  \mathrm{Id} \Big) = \nonumber \\
&&\hspace{0cm} 
= \int_{\mathrm{SO}_n {\mathbb R}} (A \cdot P) \, (A \cdot Q) \, \Big( \frac{A + A^T}{2} - \frac{1}{n} \, \mathrm{Tr} A \, \mathrm{Id} \Big) \, g(A) \, dA, \quad \forall P, \, Q \in {\mathcal A}_n. 
\label{eq:calculc4}
\end{eqnarray}
Again, taking $P$ and $Q$ as in \eqref{eq:choicePiQi} and looking at the $(p,q)$-th entry of the matrix equation \eqref{eq:calculc4}, we get
\begin{eqnarray*}
&&\hspace{-1cm}
\Big\{ \frac{1}{2} \Big( \delta_{jk} \, ( \delta_{ip} \, \delta_{\ell q} + \delta_{iq} \, \delta_{\ell p} )
+ \delta_{i \ell} \, ( \delta_{jp} \, \delta_{k q} + \delta_{jq} \, \delta_{k p} )
- \delta_{j \ell} \, ( \delta_{ip} \, \delta_{k q} + \delta_{iq} \, \delta_{k p} ) \\
&& \hspace{4cm}
- \delta_{ik} \, ( \delta_{jp} \, \delta_{\ell q} + \delta_{jq} \, \delta_{\ell p} ) \Big)
- \frac{2}{n}  \, ( \delta_{jk} \, \delta_{i \ell} - \delta_{j \ell } \, \delta_{ij} ) \, \delta_{pq} \Big\} \, C_4 \\
&& \hspace{2cm}
=\int_{\mathrm{SO}_n {\mathbb R}} (A_{ij}-A_{ji}) (A_{k\ell} - A_{\ell k}) \, \Big( \frac{A_{pq} + A_{qp}}{2} - \frac{1}{n} \, \mathrm{Tr} A \,\delta_{pq} \Big) \, g(A) \, dA. 
\end{eqnarray*}
We note that both sides of this equation are equal to $0$ if $i=j$ or $k=\ell$, so that it is valid for any integers $i, \, j, \, k, \, \ell, \, p, \, q \in \{1, \ldots, n \}$. Then, making $k=j$, $p = \ell$ and $q=i$ and summing over $i, \, j, \, \ell$, we get
$$ \frac{(n-1)(n^2-4)}{2} C_4 =  \int_{\mathrm{SO}_n {\mathbb R}} \sum_{i, \, j, \, \ell} (A_{ij}-A_{ji})  \, (A_{j \ell}-A_{\ell j}) \, \Big( \frac{A_{\ell i}+A_{i \ell}}{2} - \frac{1}{n} \, \mathrm{Tr} A \, \delta_{\ell i} \Big) \, g(A) \, dA. $$
Then, the sum inside the integral is equal to $\mathrm{Tr} A^3 - \mathrm{Tr} A \, \big( \frac{2}{n} \mathrm{Tr} A^2 - 1 \big)$, which leads to \eqref{eq:expressc4} by making $g= M_{\mathrm{Id}}$.

\medskip
\noindent
\textbf{(ii) cases $n \in \{3, \ldots, 8 \}$.} 

\noindent
\emph{(ii)-a Case $n=3$}. This is the situation studied in \cite{Degond_eal_JNLS20, Degond_etal_arxiv21, degond2017new, Degond_etal_MMS18, Degond_etal_proc19}. With $V = {\mathbb C}^3$, it is classical that $\Lambda^4(V) = \{0\}$ and we also have ${\mathbb S}_{[2,2]}(V) = \{0\}$ \cite[Exercise 19.20]{Fulton_Harris}. Furthermore, we have $\Lambda^2(V) \approx V$ through the isomorphism $\alpha$: $\Lambda^2(V) \to V$, $v \wedge w \mapsto \alpha(v \wedge w)$ such that $(\alpha(v \wedge w) \cdot z) = \det(u,v,z)$, $\forall v, \, w, \, z \in V$. In other words, $\alpha(v \wedge w) = v \times w$ is the vector product of $v$ and $w$. Through this isomorphism, \eqref{eq:defQ0Q1} becomes equivalent to \eqref{eq:decompS2}. Apart from simplifications to the computations, the stream of the proof and the final result remain identical.

\medskip
\noindent
\emph{(ii)-b Case $n = 4$.} With $V = {\mathbb C}^4$, we have $\Lambda^4 (V) = {\mathbb C} \, e_1 \wedge e_2 \wedge e_3 \wedge e_4 \approx {\mathbb C}$. Therefore, in the decomposition \eqref{eq:decompS2(Lamb)}, there are now two factors which are isomorphic to ${\mathbb C}$: $\Lambda^4 ({\mathbb C}^4)$ and $\Sigma_0$. So, when applying Schur's Lemma, we have a third term in \eqref{eq:express_tilB} and $\tilde B$ is now written: 
$$
\tilde B  = C'_3 \, {\mathcal J}_0^{-1} \circ {\mathcal Q}_0 + C'_4 \, {\mathcal J}_1^{-1} \circ {\mathcal Q}_1 + C'_5 \, {\mathcal J}_3^{-1} \circ {\mathcal T}_1,  
$$
where  the projection ${\mathcal T}_1$:  $\textrm{Sym}^2 (\Lambda^2(V)) \to \Lambda^4(V)$ is defined by \eqref{eq:tens_decomp2} and ${\mathcal J}_3$: ${\mathbb C} \, \Psi \to {\mathbb C} \, e_1 \wedge e_2 \wedge e_3 \wedge e_4$, $\lambda \, \Psi \mapsto \lambda \, e_1 \wedge e_2 \wedge e_3 \wedge e_4$ is a mapping between one-dimensional trivial representations. Introducing the automorphism $\alpha$ of $\Lambda^2(V)$ defined in Section \ref{sec:proof:expressLBi}, Case (ii), we note that ${\mathcal T}_1$ takes the form 
\begin{eqnarray*} 
{\mathcal T}_1 \big( (v_1 \wedge v_2) \circ (v_3 \wedge v_4) \big) &=& \frac{1}{3} \, \det (v_1, v_2, v_3, v_4) \, e_1 \wedge e_2 \wedge e_3 \wedge e_4 \\
&=& \frac{1}{3} \, \big( \alpha(v_1 \wedge v_2) \cdot (v_3 \wedge v_4) \big) \, e_1 \wedge e_2 \wedge e_3 \wedge e_4. 
\end{eqnarray*}
Thus, for two generic elements $P$, $Q$ of $\Lambda^2(V)$, we have 
$$ {\mathcal T}_1 (P \circ Q) = \frac{1}{3} \, \big( \alpha(P) \cdot Q \big) \, e_1 \wedge e_2 \wedge e_3 \wedge e_4. $$
Using the same methodology as in the generic case, it follows that there exist constants $C_3$, $C_4$, $C_5$ in ${\mathbb C}$ such that
\begin{equation} 
\tilde B (P \circ Q) = C_3 \, \mathrm{Tr} \, (PQ) \, \mathrm{Id} + C_4 \, \Big( \frac{PQ + QP}{2} - \frac{1}{4} \mathrm{Tr} \, (PQ) \, \mathrm{Id} \Big)  +  C_5 \, \big( \alpha(P) \cdot Q \big) \, \mathrm{Id}. 
\label{eq:express_tilB_n4}
\end{equation}
Now, we use the same methodology as in Section \ref{sec:proof:expressLBi}, Case (ii). We assume that $g$ is invariant by all outer automorphisms defined by the conjugation with an element $U \in \mathrm{O}_4{\mathbb C} \setminus \mathrm{SO}_4{\mathbb C}$. It follows that $\tilde B$ is invariant by these outer automorphisms. On the other hand, $\alpha$ is alternating by these automorphisms. Applying such an outer automorphism to \eqref{eq:express_tilB_n4}, we conclude that $C_5=0$ and the proof can be completed like in the generic case.

\medskip
\noindent
\emph{(ii)-c Cases $n = 5, \, 6, \, 7$.} We recall that for $p < n$ and $V = {\mathbb C}^n$, we have $\Lambda^p(V) \approx \Lambda^{n-p}(V)$ thanks to the isomorphism 
$$\alpha: \left\{ \begin{array}{l} \Lambda^p(V) \to \Lambda^{n-p}(V), \\ v_1 \wedge \ldots \wedge v_p \to \alpha(v_1 \wedge \ldots \wedge v_p), \end{array} \right. $$
such that
$$(\alpha(v_1 \wedge \ldots \wedge v_p) \cdot v_{p+1} \wedge \ldots \wedge v_n) = \det(v_1, \ldots , v_p, v_{p+1}, \ldots, v_n), \, \,  \forall (v_1, \ldots, v_n) \in V^n.$$ 
The inner product at the left hand side is induced by that of $V$ onto $V^{\otimes p}$ and its subspaces (such as $\Lambda^p(V)$) by $ (v_1 \otimes \ldots \otimes v_p \cdot w_1 \otimes \ldots \otimes w_p) = (v_1 \cdot w_1) \ldots (v_p \cdot w_p)$. This isomorphism is an isomorphism of representations of $\mathfrak{so}_n{\mathbb C}$. Consequently, we have $\Lambda^4(V) \approx \Lambda^1(V) = V$ if $n=5$, $\Lambda^4(V) \approx \Lambda^2(V)$ if $n=6$ and $\Lambda^4(V) \approx \Lambda^3(V)$ if $n=7$. All these spaces are irreducible representations of $\mathfrak{so}_n{\mathbb C}$ for $n=5, \, 6, \, 7$ respectively \cite[Theorem 19.2 (i) and 19.14]{Fulton_Harris}, and none of them is isomorphic to either ${\mathbb C}$ or ${\mathbb S}_{[2]}$. Thus the proof and conclusion of the generic case $n \geq 9$ apply to these three cases as well and the final result is identical. 

\medskip
\noindent
\emph{(ii)-d Case $n = 8$.} The space $\Lambda^4(V)$ with $V = {\mathbb C}^8$ is not an irreducible representation of $\mathfrak{so}_8{\mathbb C}$ but it decomposes into two non-isomorphic sub-representations $\Pi_\pm$ of equal dimensions equal to $35$ (with $\mathrm{dim} \, \Lambda^4(V) = {8 \choose 4} = 70$). The representation $\Sigma_1 \approx {\mathbb S}_{[2]}$ has also dimension $35$. However, neither $\Pi_+$ nor $\Pi_-$ is isomorphic to ${\mathbb S}_{[2]}$ (they have different heighest weight see \cite{Fulton_Harris}) so, it does not alter the fact that $\tilde B$ is given by \eqref{eq:express_tilB} and the proof can be concluded like in the generic case. This ends the proof of Lemma \ref{lem:expressLB} (ii). \endproof

\begin{remark} Following Remark \ref{rem:elementary}, we sketch a proof of Lemma \ref{lem:expressLB} (ii) relying on elementary algebra only. The proof has three steps: 
\begin{enumerate}
\item The goal is to compute $B(P,Q)\cdot S$ for $P,Q\in\mathcal{A}_n$ and $S\in \mathcal{S}_n$. By the invariance by conjugation \eqref{eq:invarB} and the spectral theorem, one can take for $S$ a diagonal matrix. 
\item By linearity, it remains to compute the quantities $B(\alpha_{ij},\alpha_{k\ell})\cdot\sigma_m$ for $i,j,k,\ell,m\in\{1,\ldots,n\}$, $i<j$, $k<\ell$ and where $\alpha_{ij} := e_i \wedge e_j$, $\sigma_m := e_m\otimes e_m$. 
\item Using the invariance by conjugation with the changes of variable described in \cite[Definition 3.1]{Degond_eal_JNLS20}, it can be proved that, at least when $n$ is large enough, $B(\alpha_{ij},\alpha_{k\ell})\cdot \sigma_m=0$ if $(i,j)\ne(k,\ell)$ and there exist $\lambda,\mu\in\mathbb{R}$ such that $B(\alpha_{ij},\alpha_{ij})\cdot \sigma_i =  B(\alpha_{ij},\alpha_{ij})\cdot \sigma_j = \lambda$ for all $i<j$ and $B(\alpha_{ij},\alpha_{ij})\cdot \sigma_m = \mu$ for all $i<j$ and $m\ne i,j$. The result then follows by a direct computation. 
\end{enumerate}
The third step is quite tedious, especially for small dimensions, and does not explicitly use the underlying algebraic structure of the problem as in the main proof presented here. Moreover, representation theory and specifically Weyl's integration formula give explicit expressions for the coefficients (see Lemma \ref{lemma:coefficientsWeyl}). Finally, unlike the approach of \cite{Degond_eal_JNLS20}, the proof presented in this article does not crucially use the fact that $\mathrm{SO}_n\mathbb{R}$ is a matrix group and may therefore be more easily generalized to other Lie groups.
\label{rem:elementary_proof}
\end{remark}

\setcounter{equation}{0}
\section{Proof of Proposition \ref{prop:alterrot} and dimension $n=3$ case}
\label{sec_comp3D}

\noindent
\textbf{Proof of Proposition \ref{prop:alterrot}} 
From \eqref{eq:orient_3}, we deduce \eqref{eq:orient_4} with ${\mathbb A}$ replaced by $\tilde {\mathbb A}$ defined by
$$\tilde {\mathbb A} \rot = - \big( (\nabla_x \wedge \Omega_1) \rot + (\Omega_1 \cdot \nabla_x ) \rot \big) . $$
We now show that $\tilde {\mathbb A} = {\mathbb A}$  with ${\mathbb A}$ given by \eqref{eq:defmathbbA}. On the one hand, we have:
\begin{eqnarray*} (\tilde {\mathbb A} \rot)_{ij} &=& -\sum_{k=1}^n \big( \partial_{x_i} (\Omega_1)_k - \partial_{x_k} (\Omega_1)_i \big)  \, \rot_{kj} - (\Omega_1 \cdot \nabla_x) \rot_{ij} \\
&=& -\sum_{k=1}^n \big( \partial_{x_i} (\Omega_1)_k - \partial_{x_k} (\Omega_1)_i \big)  \, (\Omega_j)_k - (\Omega_1 \cdot \nabla_x) (\Omega_j)_i \\
&=& - \sum_{k=1}^n (\Omega_j)_k \partial_{x_i} (\Omega_1)_k   + (\Omega_j \cdot \nabla_x) (\Omega_1)_i  - (\Omega_1 \cdot \nabla_x) (\Omega_j)_i. 
\end{eqnarray*}
On the other hand, from \eqref{eq:defmathbbA}, we have 
$${\mathbb A}_{ik} = \sum_{p,q=1}^n \Delta_{1pq} \rot_{ip} \rot_{kq}, $$
so that, using \eqref{eq:def_delta} and \eqref{eq:def_delijk}, we have
\begin{eqnarray*} 
({\mathbb A} \rot)_{ij} &=& \sum_{k=1}^n {\mathbb A}_{ik} \rot_{kj} = \sum_{k=1}^n \Delta_{1kj} \rot_{ik} = \sum_{k=1}^n \Delta_{1kj} (\Omega_k)_i \\
&=& \sum_{k=1}^n \Big( \big( (\Omega_1 \cdot \nabla_x) \Omega_k \big) \cdot \Omega_j + \big( (\Omega_k \cdot \nabla_x) \Omega_j \big) \cdot \Omega_1 + \big( (\Omega_j \cdot \nabla_x) \Omega_1 \big) \cdot \Omega_k \Big) (\Omega_k)_i. 
\end{eqnarray*}
The last term is equal to $(\Omega_j \cdot \nabla_x) (\Omega_1)_i$. For the first term, since $\Omega_k \cdot \Omega_j = \delta_{ij}$ is independent of $x$, we have 
$$ \sum_{k=1}^n  (\Omega_k)_i \, \big( (\Omega_1 \cdot \nabla_x) \Omega_k \big) \cdot \Omega_j = - \sum_{k=1}^n (\Omega_k)_i \big( (\Omega_1 \cdot \nabla_x) \Omega_j \big) \cdot \Omega_k = - (\Omega_1 \cdot \nabla_x) (\Omega_j)_i. $$
Similarly, for the second term, we have
$$ \sum_{k=1}^n  (\Omega_k)_i \, \big( (\Omega_k \cdot \nabla_x) \Omega_j \big) \cdot \Omega_1 = - \sum_{k=1}^n (\Omega_k)_i \big( (\Omega_k \cdot \nabla_x) \Omega_1 \big) \cdot \Omega_j = - \sum_{k=1}^n  (\Omega_j)_k \partial_{x_i} (\Omega_1)_k. $$
Thus, we get ${\mathbb A} \rot = \tilde {\mathbb A} \rot$ and since $\rot$ is invertible, ${\mathbb A} = \tilde {\mathbb A}$, which ends the proof. \endproof


\bigskip
\noindent
\textbf{Dimension $n=3$:} 
In \cite{Degond_etal_arxiv21}, the constants (which we will temporarily call $\tilde c_1,$ \ldots, $\tilde c_4$) were given by 
\begin{align}
\tilde c_1 &= \frac{1}{3} \, \frac{\displaystyle \int_0^{2 \pi} ( 1 + 2 \cos \theta) \, \exp \big( \kappa (1 + 2 \cos \theta) \big) \, \sin^2 \big(\frac{\theta}{2} \big) \, d\theta}{\displaystyle \int_0^{2 \pi} \exp \big( \kappa \left(1 + 2 \cos \theta \right) \big) \, \sin^2 \big(\frac{\theta}{2} \big) \, d\theta}, \label{eq:tilc1}\\
\tilde c_2 - \tilde c_4&= \frac{1}{5} \, \frac{\displaystyle \int_0^{2 \pi} (2 + 3 \cos \theta) \,  \exp \big( \kappa (1 + 2 \cos \theta) \big) \, \sin^4 \big(\frac{\theta}{2} \big) \, \cos^2 \big(\frac{\theta}{2} \big) \, d\theta}{\displaystyle  \int_0^{2 \pi} \exp \big( \kappa (1 + 2 \cos \theta) \big) \, \sin^4 \big(\frac{\theta}{2} \big) \, \cos^2 \big(\frac{\theta}{2} \big) \, d\theta}, \label{eq:tilc2}\\
\tilde c_3 &= \frac{1}{2 \kappa}, 
\label{eq:tilc3}
\\
\tilde c_4 &= \frac{1}{5} \, \frac{\displaystyle \int_0^{2 \pi}  (1 - \cos \theta) \, \exp \big( \kappa (1 + 2 \cos \theta) \big) \, \sin^4 \big(\frac{\theta}{2} \big) \, \cos^2 \big(\frac{\theta}{2} \big) \, d\theta}{\displaystyle  \int_0^{2 \pi} \exp \big( \kappa (1 + 2 \cos \theta) \big) \, \sin^4 \big(\frac{\theta}{2} \big) \, \cos^2 \big(\frac{\theta}{2} \big) \, d\theta}, \label{eq:tilc4}
\end{align}
where we have modified the expressions from \cite{Degond_etal_arxiv21} to take into account that in \cite{Degond_etal_arxiv21} the matrix inner product involved a factor $1/2$ inside the trace. This modification multiplies by a factor $2$ the expressions inside the exponentials compared with \cite{Degond_etal_arxiv21}. 
We now check that $\tilde c_i = c_i$ where $c_i$ are the constants \eqref{eq:express_c1_WIF}-\eqref{eq:express_c4_WIF} in dimension $n=3$ 

First, by comparing \eqref{eq:express_c3} and \eqref{eq:tilc3}, we see that $\tilde c_3 = c_3$. We readily notice that Eq. \eqref{eq:express_c1_WIF} for $c_1$ with $n=3$ is the same as Eq. \eqref{eq:tilc1} for $\tilde c_1$, upon changing $\sin^2 (\theta/2)$ into $(1-\cos \theta)/2$ (note that the factors $1/2$ in the numerator and denominator cancel each other). So, we have $\tilde c_1 = c_1$. In passing, we realize that the exponential factors in all the integrals for the $c$'s and $\tilde c$'s are the same, so we will only focus on the prefactors. The integrals at the denominator of Eqs. \eqref{eq:express_c2_WIF} and \eqref{eq:express_c4_WIF} for $c_2$ and $c_4$ are the same and, with $n=3$, involve the factor
\begin{equation}
\big( 1- \frac{1}{3} C_3^{(2)} \big) \, u_3 =  \big( 1 - \frac{1}{3} (1 + 2 \cos 2 \theta) \big) \, (1 - \cos \theta) = \frac{32}{3} \, \sin^4 \big(\frac{\theta}{2} \big) \, \cos^2 \big(\frac{\theta}{2} \big).  
\label{eq:checkc2c4_den}
\end{equation}
So, up to a numerical factor which we ignore for the time being, the denominators of the formulas for $c_2$ and $c_4$ match those of Eqs. \eqref{eq:tilc2} and \eqref{eq:tilc4} for $\tilde c_2$ and $\tilde c_4$. At the numerator of Eq. \eqref{eq:express_c4_WIF} for $c_4$ in the case $n=3$, we find the factor 
\begin{eqnarray} 
&&\hspace{-1cm}
\big( C_3^{(3)} - \frac{2}{3} C_3^{(1)} \, C_3^{(2)} + C_3^{(1)} \big) \, u_3 = \nonumber\\
&&\hspace{0cm}
= \big( 1 + 2 \cos 3 \theta - \frac{2}{3} (1 + 2 \cos  \theta) (1 + 2 \cos 2 \theta) + (1 + 2 \cos \theta) \big) \, (1 - \cos \theta) \nonumber \\
&&\hspace{0cm}
= \frac{8}{3} (1 - \cos \theta)^2 \, (1 - \cos^2 \theta) =  \frac{64}{3} (1 - \cos \theta) \, \sin^4 \big(\frac{\theta}{2} \big) \, \cos^2 \big(\frac{\theta}{2} \big), 
\label{eq:checkc4_num}
\end{eqnarray}
so that, still up to a numerical factor, it is equal to the numerator of Eq. \eqref{eq:tilc4} for $\tilde c_4$. Let's now check the numerical factor for $c_4$. From \eqref{eq:express_c4_WIF} and \eqref{eq:checkc2c4_den}, there is a factor $10 \times \frac{32}{3}$ at the denominator and from \eqref{eq:checkc4_num}, a factor $\frac{64}{3}$ at the numerator which results in a factor $5$ at the denominator, matching that in Eq. \eqref{eq:tilc4}. So, we have shown that $\tilde c_4 = c_4$. Finally, we need to compare $c_2$ with $\tilde c_2$ and we already know that up to a numerical factor, the denominators match. The prefactor of the exponential in the numerator of the formula for $\tilde c_2$ is $(3 + 2 \cos \theta) \, \sin^4 (\theta/2)  \, \cos^2 (\theta/2)$ and there is still a factor $5$ at the denominator. Now, at the numerator of Eq. \eqref{eq:express_c2_WIF} for $c_2$ in the case $n=3$, we find the factor 
\begin{eqnarray} 
&&\hspace{-1cm}
\big( 2 C_3^{(3)} - 3 C_3^{(1)} \, C_3^{(2)} + 7 C_3^{(1)} \big) \, u_3 = \nonumber\\
&&\hspace{-0.5cm}
= \big( 2 (1 + 2 \cos 3 \theta) - 3 (1 + 2 \cos  \theta) (1 + 2 \cos 2 \theta) + 7 (1 + 2 \cos \theta) \big) \, (1 - \cos \theta) \nonumber \\
&&\hspace{-0.5cm}
= 4 (3 + 2 \cos \theta) \, (1 - \cos^2 \theta) \, (1 - \cos \theta) =  32 (3 + 2 \cos \theta) \, \sin^4 \big(\frac{\theta}{2} \big) \, \cos^2 \big(\frac{\theta}{2} \big), 
\label{eq:checkc2_num}
\end{eqnarray}
which also coincides with the prefactor of the exponential in the numerator of the formula for $\tilde c_2$, up to a numerical constant. Concerning the numerical prefactor, there is a factor $15 \times \frac{32}{3}$ at the denominator coming from \eqref{eq:express_c2_WIF} and \eqref{eq:checkc2c4_den} and a factor $32$ at the numerator coming from \eqref{eq:checkc2_num}. So, the numerical prefactor is $\frac{1}{5}$ which corresponds to that of $\tilde c_2$. So, we have finally shown that $c_2 = \tilde c_2$, confirming that the model of \cite{Degond_etal_arxiv21} and the model found here for $n=3$ coincide, as they should.



\begin{thebibliography}{10}

\bibitem{aceves2019hydrodynamic}
P.~Aceves-S{\'a}nchez, M.~Bostan, J.-A. Carrillo, and P.~Degond.
\newblock Hydrodynamic limits for kinetic flocking models of {C}ucker-{S}male
  type.
\newblock {\em Math. Biosci. Eng.}, 16:7883--7910, 2019.

\bibitem{ahn2021emergent}
H.~Ahn, S.-Y. Ha, and W.~Shim.
\newblock Emergent dynamics of a thermodynamic {C}ucker-{S}male ensemble on
  complete {R}iemannian manifolds.
\newblock {\em Kinet. Relat. Models}, 14(2):323, 2021.

\bibitem{aoki1982simulation}
I.~Aoki.
\newblock A simulation study on the schooling mechanism in fish.
\newblock {\em Bulletin of the Japanese Society of Scientific Fisheries},
  48(8):1081--1088, 1982.

\bibitem{barbaro2016phase}
A.~B. Barbaro, J.~A. Canizo, J.~A. Carrillo, and P.~Degond.
\newblock Phase transitions in a kinetic flocking model of {C}ucker--{S}male
  type.
\newblock {\em Multiscale Model. Simul.}, 14(3):1063--1088, 2016.

\bibitem{barbaro2012phase}
A.~B. Barbaro and P.~Degond.
\newblock Phase transition and diffusion among socially interacting
  self-propelled agents.
\newblock {\em Discrete Contin. Dyn. Syst. Ser. B}, 19(3):1249--1278, 2014.

\bibitem{bazazi2008collective}
S.~Bazazi, J.~Buhl, J.~J. Hale, M.~L. Anstey, G.~A. Sword, S.~J. Simpson, and
  I.~D. Couzin.
\newblock Collective motion and cannibalism in locust migratory bands.
\newblock {\em Current Biology}, 18(10):735--739, 2008.

\bibitem{be2019statistical}
A.~Be\'er and G.~Ariel.
\newblock A statistical physics view of swarming bacteria.
\newblock {\em Movement Ecology}, 7(1):1--17, 2019.

\bibitem{bertin2006boltzmann}
E.~Bertin, M.~Droz, and G.~Gr{\'e}goire.
\newblock Boltzmann and hydrodynamic description for self-propelled particles.
\newblock {\em Phys. Rev. E}, 74(2):022101, 2006.

\bibitem{bertin2009hydrodynamic}
E.~Bertin, M.~Droz, and G.~Gr{\'e}goire.
\newblock Hydrodynamic equations for self-propelled particles: microscopic
  derivation and stability analysis.
\newblock {\em J. Phys. A}, 42(44):445001, 2009.

\bibitem{bertozzi2015ring}
A.~L. Bertozzi, T.~Kolokolnikov, H.~Sun, D.~Uminsky, and J.~Von~Brecht.
\newblock Ring patterns and their bifurcations in a nonlocal model of
  biological swarms.
\newblock {\em Commun. Math. Sci.}, 13(4):955--985, 2015.

\bibitem{bolley2012mean}
F.~Bolley, J.~A. Ca{\~n}izo, and J.~A. Carrillo.
\newblock Mean-field limit for the stochastic {V}icsek model.
\newblock {\em Appl. Math. Lett.}, 25(3):339--343, 2012.

\bibitem{bostan2013asymptotic}
M.~Bostan and J.~A. Carrillo.
\newblock Asymptotic fixed-speed reduced dynamics for kinetic equations in
  swarming.
\newblock {\em Math. Models Methods Appl. Sci.}, 23(13):2353--2393, 2013.

\bibitem{bostan2017reduced}
M.~Bostan and J.~A. Carrillo.
\newblock Reduced fluid models for self-propelled particles interacting through
  alignment.
\newblock {\em Math. Models Methods Appl. Sci.}, 27(07):1255--1299, 2017.

\bibitem{briant2020cauchy}
M.~Briant, A.~Diez, and S.~Merino-Aceituno.
\newblock Cauchy theory and mean-field limit for general {V}icsek models in
  collective dynamics.
\newblock {\em arXiv preprint arXiv:2004.00883}, 2020.

\bibitem{cao2020asymptotic}
F.~Cao, S.~Motsch, A.~Reamy, and R.~Theisen.
\newblock Asymptotic flocking for the three-zone model.
\newblock {\em Math. Biosci. Eng.}, 17(6):7692--7707, 2020.

\bibitem{carrillo2010asymptotic}
J.~A. Carrillo, M.~Fornasier, J.~Rosado, and G.~Toscani.
\newblock Asymptotic flocking dynamics for the kinetic {C}ucker--{S}male model.
\newblock {\em SIAM J. Math. Anal.}, 42(1):218--236, 2010.

\bibitem{cercignani2013mathematical}
C.~Cercignani, R.~Illner, and M.~Pulvirenti.
\newblock {\em The mathematical theory of dilute gases}, volume 106.
\newblock Springer Science \& Business Media, 2013.

\bibitem{charbonneau2013dimensional}
B.~Charbonneau, P.~Charbonneau, Y.~Jin, G.~Parisi, and F.~Zamponi.
\newblock Dimensional dependence of the {S}tokes--{E}instein relation and its
  violation.
\newblock {\em The Journal of chemical physics}, 139(16):164502, 2013.

\bibitem{chate2008collective}
H.~Chat{\'e}, F.~Ginelli, G.~Gr{\'e}goire, and F.~Raynaud.
\newblock Collective motion of self-propelled particles interacting without
  cohesion.
\newblock {\em Phys. Rev. E}, 77(4):046113, 2008.

\bibitem{clarte2019collective}
G.~Clart{\'e}, A.~Diez, and J.~Feydy.
\newblock Collective proposal distributions for nonlinear {MCMC} samplers:
  Mean-field theory and fast implementation.
\newblock {\em arXiv preprint arXiv:1909.08988}, 2019.

\bibitem{costanzo2018spontaneous}
A.~Costanzo and C.~Hemelrijk.
\newblock Spontaneous emergence of milling (vortex state) in a {V}icsek-like
  model.
\newblock {\em J. Phys. D}, 51(13):134004, 2018.

\bibitem{creppy2016symmetry}
A.~Creppy, F.~Plourabou{\'e}, O.~Praud, X.~Druart, S.~Cazin, H.~Yu, and
  P.~Degond.
\newblock Symmetry-breaking phase transitions in highly concentrated semen.
\newblock {\em Journal of The Royal Society Interface}, 13(123):20160575, 2016.

\bibitem{cucker2007emergent}
F.~Cucker and S.~Smale.
\newblock Emergent behavior in flocks.
\newblock {\em IEEE Trans. Automat. Control}, 52(5):852--862, 2007.

\bibitem{Degond_eal_JNLS20}
P.~Degond, A.~Diez, A.~Frouvelle, and S.~Merino-Aceituno.
\newblock Phase transitions and macroscopic limits in a {B}{G}{K} model of
  body-attitude coordination.
\newblock {\em J. Nonlinear Sci.}, 30:2671--2736, 2020.

\bibitem{Degond_etal_inprep}
P.~Degond, A.~Diez, A.~Frouvelle, S.~Merino-Aceituno, and A.~Trescases.
\newblock Qualitative properties of continuum models of body attitude
  coordination.
\newblock {\em in preparation}, 2022.

\bibitem{Degond_etal_arxiv21}
P.~Degond, A.~Diez, and M.~Na.
\newblock Bulk topological states in a new collective dynamics model.
\newblock {\em arXiv preprint arXiv:2101.10864}, 2021.

\bibitem{degond2013macroscopic}
P.~Degond, A.~Frouvelle, and J.-G. Liu.
\newblock Macroscopic limits and phase transition in a system of self-propelled
  particles.
\newblock {\em J. Nonlinear Sci.}, 23(3):427--456, 2013.

\bibitem{degond2015phase}
P.~Degond, A.~Frouvelle, and J.-G. Liu.
\newblock Phase transitions, hysteresis, and hyperbolicity for self-organized
  alignment dynamics.
\newblock {\em Arch. Ration. Mech. Anal.}, 216(1):63--115, 2015.

\bibitem{degond2017new}
P.~Degond, A.~Frouvelle, and S.~Merino-Aceituno.
\newblock A new flocking model through body attitude coordination.
\newblock {\em Math. Models Methods Appl. Sci.}, 27(06):1005--1049, 2017.

\bibitem{Degond_etal_proc19}
P.~Degond, A.~Frouvelle, S.~Merino-Aceituno, and A.~Trescases.
\newblock Alignment of self-propelled rigid bodies: from particle systems to
  macroscopic equations.
\newblock In {\em International workshop on Stochastic Dynamics out of
  Equilibrium}, pages 28--66. Springer, 2017.

\bibitem{Degond_etal_MMS18}
P.~Degond, A.~Frouvelle, S.~Merino-Aceituno, and A.~Trescases.
\newblock Quaternions in collective dynamics.
\newblock {\em Multiscale Model. Simul.}, 16(1):28--77, 2018.

\bibitem{degond2008continuum}
P.~Degond and S.~Motsch.
\newblock Continuum limit of self-driven particles with orientation
  interaction.
\newblock {\em Math. Models Methods Appl. Sci.}, 18(supp01):1193--1215, 2008.

\bibitem{diez2020propagation}
A.~Diez.
\newblock Propagation of chaos and moderate interaction for a piecewise
  deterministic system of geometrically enriched particles.
\newblock {\em Electron. J. Probab.}, 25:1--38, 2020.

\bibitem{diez2021sisyphe}
A.~Diez.
\newblock {S}i{S}y{P}{H}{E}: A {P}ython package for the simulation of systems
  of interacting mean-field particles with high efficiency.
\newblock {\em Journal of Open Source Software}, 6(65):3653, 2021.

\bibitem{fetecau2021emergent}
R.~C. Fetecau, S.-Y. Ha, and H.~Park.
\newblock Emergent behaviors of rotation matrix flocks.
\newblock {\em arXiv preprint arXiv:2103.06458}, 2021.

\bibitem{fetecau2021well}
R.~C. Fetecau and F.~S. Patacchini.
\newblock Well-posedness of an interaction model on {R}iemannian manifolds.
\newblock {\em arXiv preprint arXiv:2109.03959}, 2021.

\bibitem{fetecau2019self}
R.~C. Fetecau and B.~Zhang.
\newblock Self-organization on {R}iemannian manifolds.
\newblock {\em J. Geom. Mech.}, 11(3):397--426, 2019.

\bibitem{figalli2018global}
A.~Figalli, M.-J. Kang, and J.~Morales.
\newblock Global well-posedness of the spatially homogeneous
  {K}olmogorov--{V}icsek model as a gradient flow.
\newblock {\em Arch. Ration. Mech. Anal.}, 227(3):869--896, 2018.

\bibitem{frouvelle2012continuum}
A.~Frouvelle.
\newblock A continuum model for alignment of self-propelled particles with
  anisotropy and density-dependent parameters.
\newblock {\em Math. Models Methods Appl. Sci.}, 22(07):1250011, 2012.

\bibitem{frouvelle2020body}
A.~Frouvelle.
\newblock Body-attitude alignment: first order phase transition, link with
  rodlike polymers through quaternions, and stability.
\newblock {\em arXiv preprint arXiv:2011.14891}, 2020.

\bibitem{frouvelle2012dynamics}
A.~Frouvelle and J.-G. Liu.
\newblock Dynamics in a kinetic model of oriented particles with phase
  transition.
\newblock {\em SIAM J. Math. Anal.}, 44(2):791--826, 2012.

\bibitem{Fulton_Harris}
W.~Fulton and J.~Harris.
\newblock {\em Representation theory: a first course}, volume 129.
\newblock Springer Science \& Business Media, 2013.

\bibitem{gamba2016global}
I.~M. Gamba and M.-J. Kang.
\newblock Global weak solutions for {K}olmogorov--{V}icsek type equations with
  orientational interactions.
\newblock {\em Arch. Ration. Mech. Anal.}, 222(1):317--342, 2016.

\bibitem{giniunaite2020modelling}
R.~Gini{\=u}nait{\.e}, R.~E. Baker, P.~M. Kulesa, and P.~K. Maini.
\newblock Modelling collective cell migration: neural crest as a model
  paradigm.
\newblock {\em J. Math. Biol.}, 80(1):481--504, 2020.

\bibitem{golse2019mean}
F.~Golse and S.-Y. Ha.
\newblock A mean-field limit of the {L}ohe matrix model and emergent dynamics.
\newblock {\em Arch. Ration. Mech. Anal.}, 234(3):1445--1491, 2019.

\bibitem{griette2019kinetic}
Q.~Griette and S.~Motsch.
\newblock Kinetic equations and self-organized band formations.
\newblock In {\em Active Particles, Volume 2}, pages 173--199. Springer, 2019.

\bibitem{ha2017emergent}
S.-Y. Ha, D.~Ko, and S.~W. Ryoo.
\newblock Emergent dynamics of a generalized {L}ohe model on some class of
  {L}ie groups.
\newblock {\em J. Stat. Phys.}, 168(1):171--207, 2017.

\bibitem{ha2009simple}
S.-Y. Ha and J.-G. Liu.
\newblock A simple proof of the {C}ucker-{S}male flocking dynamics and
  mean-field limit.
\newblock {\em Commun. Math. Sci.}, 7(2):297--325, 2009.

\bibitem{ha2008particle}
S.-Y. Ha and E.~Tadmor.
\newblock From particle to kinetic and hydrodynamic descriptions of flocking.
\newblock {\em arXiv preprint arXiv:0806.2182}, 2008.

\bibitem{haskovec2021simple}
J.~Haskovec.
\newblock A simple proof of asymptotic consensus in the {H}egselmann--{K}rause
  and {C}ucker--{S}male models with normalization and delay.
\newblock {\em SIAM J. Appl. Dyn. Syst.}, 20(1):130--148, 2021.

\bibitem{hildenbrandt2010self}
H.~Hildenbrandt, C.~Carere, and C.~K. Hemelrijk.
\newblock Self-organized aerial displays of thousands of starlings: a model.
\newblock {\em Behavioral Ecology}, 21(6):1349--1359, 2010.

\bibitem{jiang2016hydrodynamic}
N.~Jiang, L.~Xiong, and T.-F. Zhang.
\newblock Hydrodynamic limits of the kinetic self-organized models.
\newblock {\em SIAM J. Math. Anal.}, 48(5):3383--3411, 2016.

\bibitem{lopez2012behavioural}
U.~Lopez, J.~Gautrais, I.~D. Couzin, and G.~Theraulaz.
\newblock From behavioural analyses to models of collective motion in fish
  schools.
\newblock {\em Interface focus}, 2(6):693--707, 2012.

\bibitem{motsch2011new}
S.~Motsch and E.~Tadmor.
\newblock A new model for self-organized dynamics and its flocking behavior.
\newblock {\em J. Stat. Phys.}, 144(5):923--947, 2011.

\bibitem{pinnau2017consensus}
R.~Pinnau, C.~Totzeck, O.~Tse, and S.~Martin.
\newblock A consensus-based model for global optimization and its mean-field
  limit.
\newblock {\em Math. Models Methods Appl. Sci.}, 27(01):183--204, 2017.

\bibitem{saito2017understanding}
A.~Saito, T.~I. Farhana, A.~M.~R. Kabir, D.~Inoue, A.~Konagaya, K.~Sada, and
  A.~Kakugo.
\newblock Understanding the emergence of collective motion of microtubules
  driven by kinesins: role of concentration of microtubules and depletion
  force.
\newblock {\em RSC advances}, 7(22):13191--13197, 2017.

\bibitem{sarlette2010coordinated}
A.~Sarlette, S.~Bonnabel, and R.~Sepulchre.
\newblock Coordinated motion design on lie groups.
\newblock {\em IEEE Trans. Automat. Control}, 55(5):1047--1058, 2010.

\bibitem{sarlette2009consensus}
A.~Sarlette and R.~Sepulchre.
\newblock Consensus optimization on manifolds.
\newblock {\em SIAM J. Control Optim.}, 48(1):56--76, 2009.

\bibitem{sarlette2009autonomous}
A.~Sarlette, R.~Sepulchre, and N.~E. Leonard.
\newblock Autonomous rigid body attitude synchronization.
\newblock {\em Automatica J. IFAC}, 45(2):572--577, 2009.

\bibitem{sepulchre2010consensus}
R.~Sepulchre, A.~Sarlette, and P.~Rouchon.
\newblock Consensus in non-commutative spaces.
\newblock In {\em 49th IEEE conference on decision and control (CDC)}, pages
  6596--6601. IEEE, 2010.

\bibitem{shellard2020rules}
A.~Shellard and R.~Mayor.
\newblock Rules of collective migration: from the wildebeest to the neural
  crest.
\newblock {\em Philos. Trans. Roy. Soc. B}, 375(1807):20190387, 2020.

\bibitem{Simon}
B.~Simon.
\newblock {\em Representations of finite and compact groups}.
\newblock Number~10 in Graduate Studies in Mathematics. American Mathematical
  Soc., 1996.

\bibitem{toner1998flocks}
J.~Toner and Y.~Tu.
\newblock Flocks, herds, and schools: A quantitative theory of flocking.
\newblock {\em Phys. Rev. E}, 58(4):4828, 1998.

\bibitem{vicsek1995novel}
T.~Vicsek, A.~Czir{\'o}k, E.~Ben-Jacob, I.~Cohen, and O.~Shochet.
\newblock Novel type of phase transition in a system of self-driven particles.
\newblock {\em Phys. Rev. Lett.}, 75(6):1226, 1995.

\bibitem{warren2018collective}
W.~H. Warren.
\newblock Collective motion in human crowds.
\newblock {\em Current directions in psychological science}, 27(4):232--240,
  2018.

\bibitem{zhang2017local}
T.-F. Zhang and N.~Jiang.
\newblock A local existence of viscous self-organized hydrodynamic model.
\newblock {\em Nonlinear Anal. Real World Appl.}, 34:495--506, 2017.

\end{thebibliography}

\end{document}